\documentstyle[12pt,axodraw,graphicx]{article}
\begin{document}
\baselineskip 0.6cm

\def\simgt{\mathrel{\lower2.5pt\vbox{\lineskip=0pt\baselineskip=0pt
           \hbox{$>$}\hbox{$\sim$}}}}
\def\simlt{\mathrel{\lower2.5pt\vbox{\lineskip=0pt\baselineskip=0pt
           \hbox{$<$}\hbox{$\sim$}}}}
\def\npropto{\propto\!\!\!\!\!\not\,\,\,\,\,}

\begin{titlepage}

\begin{flushright}
UCB-PTH-08/64
\end{flushright}

\vskip 2.4cm

\begin{center}

{\Large \bf 
Naturally Flavorful Supersymmetry at the LHC
}

\vskip 1.0cm

{\large Yasunori Nomura and Daniel Stolarski}

\vskip 0.4cm

{\it Department of Physics, University of California,
     Berkeley, CA 94720} \\
{\it Theoretical Physics Group, Lawrence Berkeley National Laboratory,
     Berkeley, CA 94720}

\vskip 1.2cm

\abstract{The suppression of flavor and $CP$ violation in supersymmetric 
 theories may be due to the mechanism responsible for the structure 
 of the Yukawa couplings.  We study model independently the compatibility 
 between low energy flavor and $CP$ constraints and observability of 
 superparticles at the LHC, assuming a generic correlation between the 
 Yukawa couplings and the supersymmetry breaking parameters.  We find 
 that the superpotential operators that generate scalar trilinear 
 interactions are generically problematic.  We discuss several ways 
 in which this tension is naturally avoided.  In particular, we focus 
 on several frameworks in which the dangerous operators are naturally 
 absent.  These frameworks can be combined with many theories of flavor, 
 including those with (flat or warped) extra dimensions, strong dynamics, 
 or flavor symmetries.  We show that the resulting theories can avoid 
 all the low energy constraints while keeping the superparticles light. 
 The intergenerational mass splittings among the sfermions can reflect 
 the structure of the underlying flavor theory, and can be large enough 
 to be measurable at the LHC.  Detailed observations of the superparticle 
 spectrum may thus provide new handles on the origin of the flavor 
 structure of the standard model.}

\end{center}
\end{titlepage}

\section{Introduction}
\label{sec:intro}

The origin of the flavor structure is one of the deepest mysteries of 
the standard model.  In the absence of the Yukawa couplings (and neutrino 
masses), the standard model respects the following flavor symmetry:%
\footnote{$G_{\rm flavor}^{\rm SM}$ contains hypercharge, baryon 
 number, and lepton number.  Out of the five $U(1)$ factors, only 
 hypercharge and baryon minus lepton number are anomaly free with 
 respect to the standard model gauge group.}
\begin{equation}
  G_{\rm flavor}^{\rm SM} \equiv U(3)^5 
  = U(3)_Q \times U(3)_U \times U(3)_D \times U(3)_L \times U(3)_E.
\label{eq:G_flavor}
\end{equation}
Is this symmetry a mere artifact of the low-energy Lagrangian, or 
is it (or its subgroup) physically realized at high energies and 
spontaneously broken to produce the Yukawa couplings?  If the latter, 
what is the fundamental flavor group $G_{\rm flavor}$ ($\subset 
G_{\rm flavor}^{\rm SM}$), and how is it broken?  In the standard 
model, these questions can be explored only through the observed pattern 
of masses and mixings of the quarks and leptons, making it difficult 
to arrive at conclusive answers.

Theories beyond the standard model may provide additional clues to 
address the puzzle of flavor, since the new physics sector may contain 
new information on flavor.  In supersymmetric theories, for example, 
supersymmetry breaking parameters for squarks and sleptons may carry 
such information.  It is, however, not obvious how much new information 
one can expect.  In supersymmetric theories, generic weak scale 
values for the supersymmetry breaking parameters
\begin{equation}
  (m_\Phi^2)_{ij} \sim m_{\rm SUSY}^2,
\qquad
  (a_f)_{ij} \sim m_{\rm SUSY},
\label{eq:generic-SUSY}
\end{equation}
lead to flavor changing neutral currents far in excess of current 
experimental bounds.  Here, $(m_\Phi^2)_{ij}$ ($\Phi=Q,U,D,L,E$) and 
$(a_f)_{ij}$ ($f=u,d,e$) represent scalar squared masses and trilinear 
interactions, respectively, $i,j=1,2,3$ are generation indices, and 
$m_{\rm SUSY}$ is a parameter of order the weak scale.  A common 
solution to this problem is to assume flavor universality
\begin{equation}
  (m_\Phi^2)_{ij} \propto \delta_{ij} \,\,{\rm or} \ll m_{\rm SUSY}^2,
\qquad
  (a_f)_{ij} \propto (y_f)_{ij} \,\,{\rm or} \ll m_{\rm SUSY},
\label{eq:univ-SUSY}
\end{equation}
at a scale $M$ where the supersymmetry breaking parameters are 
generated~[\ref{Dimopoulos:1981zb:X}~--~\ref{Kaplan:1999ac:X}]. 
Here, $(y_f)_{ij}$ are the Yukawa matrices.  This assumption, however, 
greatly reduces the flavor information encoded in low energy supersymmetry. 
A nontrivial flavor structure can still be found in the low energy 
squark and slepton masses due to renormalization group evolution 
below $M$.  This structure, however, does not carry any information 
on flavor beyond Yukawa couplings $\Phi$ have, although it does allow 
us to explore some of these couplings that cannot be probed in the 
standard model~\cite{Hall:1985dx}.  In addition, the size of the relevant 
flavor nonuniversality is typically so small that most of the interesting 
parameters can be probed only indirectly through low energy flavor 
and $CP$ violating processes.

In this paper we study the question: is there a natural and generic 
framework for supersymmetry which is sufficiently ``flavorful,'' i.e. 
which allows us to obtain more detailed information on flavor through 
measurements of superparticle masses and interactions at the LHC? 
This is not trivial because such a framework must satisfy stringent 
constraints from flavor and $CP$ violation while the deviation from 
universality must be sufficiently large to be experimentally observable. 
In particular, in order for the flavor information to be extracted 
at the LHC, superparticles must be light enough to be produced at 
the LHC, making it more difficult to satisfy the bounds from the 
low energy flavor and $CP$ violating processes.

While it is not too difficult to consider an ad hoc deviation from 
universality that is measurable at the LHC and not excluded by the 
low energy data, an important question is if there is a theoretically 
well-motivated setup which naturally produces measurable effects that 
are consistent with the low energy experiments and encode information 
on the origin of flavor.  In a previous paper with Papucci, we 
studied a simple setup in which flavor changing interactions 
in the supersymmetry breaking parameters are scaled by factors 
associated with the Yukawa couplings~\cite{Nomura:2007ap}.  We 
showed that such a setup can avoid all the low energy constraints, 
while giving interesting flavor signatures at the LHC.  This clearly 
illustrates that there is an interesting, natural stage between 
Eqs.~(\ref{eq:generic-SUSY}, \ref{eq:univ-SUSY}).  In fact, there 
have been many models proposed to address the problem of flavor 
changing neutral currents, in which flavor violation in the 
supersymmetry breaking parameters is somehow related to the Yukawa 
couplings~[\ref{Dine:1993np:X}~--~\ref{Choi:2008hn:X}].  While many of 
these models require rather special structures or setups to avoid all 
the current experimental bounds, the analysis of Ref.~\cite{Nomura:2007ap} 
suggests that the minimal structure needed to obtain a consistent 
framework for flavor signatures at the LHC may, in fact, be much 
simpler.

In this paper we study the tension between LHC observability and 
constraints from low energy flavor and $CP$ violation in generic 
supersymmetric theories in which the structure of the supersymmetry 
breaking parameters is correlated with that of the Yukawa couplings. 
An interesting general point emphasized in Ref.~\cite{Nomura:2007ap} 
(see also~\cite{Kitano:2006ws}) is that among the operators giving 
supersymmetry breaking parameters, a class of operators in the 
superpotential
\begin{equation}
  W \sim X Q_i U_j H_u,\, X Q_i D_j H_d,\, X L_i E_j H_d,
\label{eq:A-op}
\end{equation}
generally leads to a strong tension.  Here, $X$ represents a chiral 
superfield whose $F$-term vacuum expectation value (VEV) is responsible 
for supersymmetry breaking.  We first elucidate this point, and define 
what we call the superpotential flavor problem.  We emphasize that 
the problem is general and does not depend on any particular theory 
of flavor.  In fact, this problem is part of a more general problem 
associated with left-right propagation of sfermions, which is also 
discussed in detail.

We then discuss how the problems described above can be solved.  We 
present several possibilities that can avoid the stringent constraints 
from left-right sfermion propagation without suppressing the operators 
of Eq.~(\ref{eq:A-op}).  We also present simple frameworks in which the 
operators of Eq.~(\ref{eq:A-op}) are naturally absent.  These frameworks 
can be combined with many theories of flavor, including theories with 
(flat or warped) extra dimensions, strong dynamics, or flavor symmetries. 
We perform detailed studies of the constraints from low energy flavor 
and $CP$ violation within these frameworks, and find that they can 
naturally avoid all the constraints while preserving the observability 
of superparticles at the LHC.  We also find that the intergenerational 
mass splittings among sfermions can show a variety of patterns depending 
on the details of the underlying flavor theory, allowing us to gain 
additional handles on the origin of flavor at the LHC.

The organization of the paper is as follows.  In section~\ref{sec:problem} 
we discuss constraints from low energy flavor and $CP$ violation 
arising from left-right propagation of sfermions.  We emphasize the 
model-independent nature of the problem associated with the operators 
of Eq.~(\ref{eq:A-op}), but also discuss additional stringent model 
dependent constraints.  In section~\ref{sec:solutions} we discuss 
possibilities to avoid these constraints.  In particular, we present 
simple frameworks in which the operators of Eq.~(\ref{eq:A-op}) are 
naturally absent.  In section~\ref{sec:spectra} we study constraints 
from flavor and $CP$ violation, including ones arising from left-left 
and right-right sfermion propagation, in these frameworks.  We also 
analyze the size of the intergenerational mass splittings among sfermions, 
and find that they can differ significantly from the flavor universal 
case.  Prospects for observing these features at the LHC are discussed 
in section~\ref{sec:LHC}.  Finally, discussion and conclusions are 
given in section~\ref{sec:concl}.

\section{The Supersymmetric Left-Right Flavor Problem}
\label{sec:problem}

The flavor problem in supersymmetric models is typically phrased such 
that generic supersymmetry breaking parameters, Eq.~(\ref{eq:generic-SUSY}), 
lead to excessive flavor and $CP$ violation at low energies.  This, 
however, neglects the possibility that the physics responsible for the 
observed Yukawa couplings also controls the pattern of the supersymmetry 
breaking parameters.  Here we argue that there is a {\it generic} tension 
between weak scale supersymmetry and low energy flavor and $CP$ violation 
even if we take this possibility into account.  Throughout the discussion, 
we assume that $CP$ violating effects not associated with flavor, 
e.g. those arising from a nontrivial phase in the Higgs sector, are 
adequately suppressed.  We also assume that the strong $CP$ problem 
is solved.

\subsection{Flavor (non)universality in the operator language}
\label{subsec:gen-ops}

We begin our discussion by listing all the operators in the supersymmetric 
standard model (SSM).  In the gauge sector, the relevant operators are
\begin{equation}
  {\cal O}_{g_A}:\, 
    \int\!d^2\theta\, \frac{1}{4 g_A^2} 
      {\cal W}^{A\alpha} {\cal W}^A_\alpha + {\rm h.c.},
\qquad\quad
  {\cal O}_{\lambda_A}:\, 
    \int\!d^2\theta\, \eta_A \frac{X}{M} 
      {\cal W}^{A\alpha} {\cal W}^A_\alpha + {\rm h.c.},
\label{eq:gauge-sector}
\end{equation}
where $A=1,2,3$ represents the standard model gauge group, $U(1)_Y$, 
$SU(2)_L$ and $SU(3)_C$, and ${\cal O}_{g_A}$ and ${\cal O}_{\lambda_A}$ 
give the gauge kinetic terms and the gaugino masses, respectively.%
\footnote{We define an operator ${\cal O}$ to be the entire term that 
 appears in the Lagrangian, including the coefficient.}
Here, $X$ is the supersymmetry breaking superfield, $\langle X \rangle 
= \theta^2 F_X$, and $M$ characterizes a scale at which supersymmetry 
breaking effects are mediated to the SSM sector.  Since $F_X/M$ 
sets the scale for superparticle masses, we consider $F_X/M \approx 
O({\rm TeV})$.  In the minimal supersymmetric standard model (MSSM), 
the Higgs sector operators are given by
\begin{equation}
  {\cal O}_{Z_H}:\, 
    \int\!d^4\theta\, Z_H\, H^\dagger H,
\label{eq:Higgs-sector-1}
\end{equation}
\begin{equation}
  {\cal O}_{\kappa_H}:\, 
    \int\!d^4\theta\, \kappa_H \frac{X^\dagger X}{M^2} H^\dagger H,
\qquad\quad
  {\cal O}_{\eta_H}:\, 
    \int\!d^4\theta\, \eta_H \frac{X}{M} H^\dagger H + {\rm h.c.},
\label{eq:Higgs-sector-2}
\end{equation}
\begin{equation}
  {\cal O}_{\mu}:\, 
    \int\!d^4\theta\, \eta_\mu \frac{X^\dagger}{M} H_u H_d + {\rm h.c.},
\qquad\quad
  {\cal O}_{b}:\, 
    \int\!d^4\theta\, \kappa_b \frac{X^\dagger X}{M^2} 
      H_u H_d + {\rm h.c.},
\label{eq:Higgs-sector-3}
\end{equation}
\begin{equation}
  {\cal O}_{\rm SUGRA}:\, 
    \int\!d^4\theta\, \lambda_H\, H_u H_d + {\rm h.c.},
\label{eq:Higgs-sector-4}
\end{equation}
where $H = H_u, H_d$.  The operators ${\cal O}_{Z_H}$ give the kinetic 
terms, while the rest provide the supersymmetric mass, $\mu$, and the 
holomorphic and non-holomorphic supersymmetry breaking squared masses, 
$b$, $m_{H_u}^2$ and $m_{H_d}^2$, for the Higgs doublets.%
\footnote{Here we have neglected the tree-level superpotential operator 
 $\int\!d^2\theta\, \mu_0\, H_u H_d + {\rm h.c.}$.  In order to have 
 weak scale values for $\mu$ and $b$, the coefficient of this operator 
 must be suppressed: $\mu_0 \simlt O({\rm TeV})$.}
The last operator is relevant only in the context of 
supergravity.  In non-minimal models, e.g. in models with 
extra gauge groups and/or singlet fields, the set of operators 
in Eqs.~(\ref{eq:gauge-sector}~--~\ref{eq:Higgs-sector-4}) 
is extended.

The operators described above (or their extensions in non-minimal models) 
do not introduce flavor violation.  Flavor violation may arise when we 
introduce matter fields.  With matter fields, we can write operators
\begin{equation}
  {\cal O}_{Z_\Phi}:\, 
    \int\!d^4\theta\, (Z_\Phi)_{ij} \Phi_i^\dagger \Phi_j,
\label{eq:matter-sector-1}
\end{equation}
\begin{equation}
  {\cal O}_{\kappa_\Phi}:\, 
    \int\!d^4\theta\, (\kappa_\Phi)_{ij} 
      \frac{X^\dagger X}{M^2} \Phi_i^\dagger \Phi_j,
\qquad\quad
  {\cal O}_{\eta_\Phi}:\, 
    \int\!d^4\theta\, (\eta_\Phi)_{ij} 
      \frac{X}{M} \Phi_i^\dagger \Phi_j + {\rm h.c.},
\label{eq:matter-sector-2}
\end{equation}
\begin{equation}
  {\cal O}_{\lambda_f}:\, 
    \int\!d^2\theta\, (\lambda_f)_{ij} \Phi_{Li} \Phi_{Rj} H + {\rm h.c.},
\qquad\quad
  {\cal O}_{\zeta_f}:\, 
    \int\!d^2\theta\, (\zeta_f)_{ij} \frac{X}{M} 
      \Phi_{Li} \Phi_{Rj} H + {\rm h.c.},
\label{eq:Yukawa-sector}
\end{equation}
where $\Phi = \Phi_L, \Phi_R$ with $\Phi_L = Q,L$ and $\Phi_R = U,D,E$ 
represents matter fields, $i,j = 1,2,3$ are generation indices, and 
$f=u,d,e$ corresponds to $\{ \Phi_L, \Phi_R, H \} = \{ Q, U, H_u \}, 
\{ Q, D, H_d \}, \{ L, E, H_d \}$.  Here, $Z_\Phi$ and $\kappa_\Phi$ 
are $3 \times 3$ Hermitian matrices, while $\eta_\Phi$, $\lambda_f$ 
and $\zeta_f$ are general complex $3 \times 3$ matrices.  The operators 
${\cal O}_{Z_\Phi}$ give the kinetic terms, ${\cal O}_{\lambda_f}$ the 
Yukawa couplings, and ${\cal O}_{\kappa_\Phi}$, ${\cal O}_{\eta_\Phi}$ 
and ${\cal O}_{\zeta_f}$ the soft supersymmetry breaking parameters.

Flavor universality is the assumption that
\begin{equation}
  (\kappa_\Phi)_{ij} \propto (Z_\Phi)_{ij},
\qquad\quad
  (\eta_\Phi)_{ij} \propto (Z_\Phi)_{ij},
\qquad\quad
  (\zeta_f)_{ij} \propto (\lambda_f)_{ij},
\label{eq:univ-1}
\end{equation}
for all $\Phi = Q,U,D,L,E$ and $f = u,d,e$, which leads to supersymmetry 
breaking parameters of the form
\begin{equation}
  (m_\Phi^2)_{ij} \propto \delta_{ij},
\qquad\quad
  (a_f)_{ij} \propto (y_f)_{ij},
\label{eq:univ-2}
\end{equation}
where $(m_\Phi^2)_{ij}$, $(a_f)_{ij}$ and $(y_f)_{ij}$ represent the 
scalar squared masses, scalar trilinear interactions, and the Yukawa 
couplings in the basis where the fields are canonically normalized. 
In fact, the three conditions of Eq.~(\ref{eq:univ-1}) could each be 
replaced by
\begin{equation}
  |(\kappa_\Phi)_{ij}| \ll |\eta_A|^2,
\qquad\quad
  |(\eta_\Phi)_{ij}| \ll |\eta_A|,
\qquad\quad
  |(\zeta_f)_{ij}| \ll |(\lambda_f)_{ij} \eta_A|,
\label{eq:univ-3}
\end{equation}
since then the low energy supersymmetry breaking parameters, generated 
by SSM renormalization group evolution, take approximately the form 
of Eq.~(\ref{eq:univ-2}).

Deviations from Eqs.~(\ref{eq:univ-1}, \ref{eq:univ-3}) generically
lead to flavor and $CP$ violating effects.  If the supersymmetry breaking 
parameters take the flavor universal form at some scale below the scale 
of flavor physics $M_F$, then small deviations from universality are 
caused only by SSM renormalization group evolution below that scale, 
which do not provide much insight into the origin of flavor at the LHC. 
On the other hand, if $M_F \simlt M$ (or if the mediation mechanism of 
supersymmetry breaking somehow carries information on physics responsible 
for the Yukawa structure), then we expect that the supersymmetry 
breaking parameters have an intrinsic flavor nonuniversality, which 
contains information on the physics of flavor at $M_F$.  Of course, 
this deviation from universality cannot be arbitrary.  In order to 
satisfy all the constraints while keeping superparticles within the 
reach of the LHC, the deviation must somehow be correlated with the 
Yukawa structure.  This is, however, precisely what we expect if the 
supersymmetry breaking parameters feel the physics responsible for 
the flavor structure.

\subsection{Generic scalar trilinear interactions}
\label{subsec:A-terms}

We consider the case where $M_F \simlt M$ and flavor nonuniversality in 
the operators of Eqs.~(\ref{eq:matter-sector-1}~--~\ref{eq:Yukawa-sector}) 
at $M_F$ is controlled by the physics responsible for the structure 
of the Yukawa couplings.  This provides a possibility of avoiding 
the low energy constraints without imposing flavor universality, 
allowing us to probe the origin of flavor through the superparticle 
spectrum.  In general, correlations between the structure of the 
Yukawa couplings and that of the nonuniversality in the operators of 
Eqs.~(\ref{eq:matter-sector-1}~--~\ref{eq:Yukawa-sector}) are model 
dependent.  However, one class of operators, ${\cal O}_{\zeta_f}$ 
in Eq.~(\ref{eq:Yukawa-sector}), is expected to have a structure 
similar to the Yukawa couplings.  This is relatively model independent 
because the matter and Higgs fields appear in ${\cal O}_{\zeta_f}$ 
in precisely the same way as in ${\cal O}_{\lambda_f}$, which 
produces the Yukawa couplings.

Suppose that the Yukawa couplings
\begin{equation}
  (y_f)_{ij} = (\lambda_f)_{kl}\, \Bigl( Z_{\Phi_L}^{-1/2} \Bigr)_{ki} 
    \Bigl( Z_{\Phi_R}^{-1/2} \Bigr)_{lj} Z_H^{-1/2},
\label{eq:Yukawa}
\end{equation}
have a hierarchical structure as a result of some flavor physics, 
for example, physics associated with spontaneous breaking of a flavor 
symmetry or wavefunction profiles of matter fields in extra dimensions. 
We then expect that the scalar trilinear interactions generated by 
${\cal O}_{\zeta_f}$,
\begin{equation}
  (a_f)_{ij} = -(\zeta_f)_{kl}\, \Bigl( Z_{\Phi_L}^{-1/2} \Bigr)_{ki} 
    \Bigl( Z_{\Phi_R}^{-1/2} \Bigr)_{lj} Z_H^{-1/2}\, \frac{F_X}{M},
\label{eq:trilinear}
\end{equation}
also have a similar structure.  Specifically, we can consider that the 
Yukawa couplings take the form
\begin{equation}
  (y_u)_{ij} \approx {\cal E}^u_{ij}\, \tilde{y},
\qquad
  (y_d)_{ij} \approx {\cal E}^d_{ij}\, \tilde{y},
\qquad
  (y_e)_{ij} \approx {\cal E}^e_{ij}\, \tilde{y},
\label{eq:y_f-E}
\end{equation}
and that the observed structure for the quark and lepton masses and 
mixings is generated by the ``suppression factors'' ${\cal E}^f_{ij}$. 
Here, $\tilde{y}$ represents the ``natural'' size of the couplings 
before taking into account the origin of the flavor structure; for 
example, we expect $\tilde{y} \approx O(1)$ if the relevant physics 
is weakly coupled, but it could be as large as of $O(4\pi)$ if strongly 
coupled.  The scalar trilinear interactions are then expected to take 
the form
\begin{equation}
  (a_u)_{ij} \approx {\cal E}^u_{ij} \frac{\tilde{\zeta} F_X}{M},
\qquad
  (a_d)_{ij} \approx {\cal E}^d_{ij} \frac{\tilde{\zeta} F_X}{M},
\qquad
  (a_e)_{ij} \approx {\cal E}^e_{ij} \frac{\tilde{\zeta} F_X}{M},
\label{eq:a_f-E}
\end{equation}
where $\tilde{\zeta}$ again represents the ``natural'' size of the 
coefficients.  Note that $O(1)$ coefficients are omitted in the 
expressions of Eqs.~(\ref{eq:y_f-E}, \ref{eq:a_f-E}); for example, 
we expect that $(a_f)_{ij}$ is in general not proportional to 
$(y_f)_{ij}$ because of an arbitrary $O(1)$ coefficient in each 
element.

The structure of Eqs.~(\ref{eq:y_f-E}, \ref{eq:a_f-E}) is expected 
to appear in most theories of flavor.  A special case is when 
${\cal E}^f_{ij}$ factorize as ${\cal E}^u_{ij} = \epsilon_{Q_i} 
\epsilon_{U_j}$, ${\cal E}^d_{ij} = \epsilon_{Q_i} \epsilon_{D_j}$ 
and ${\cal E}^e_{ij} = \epsilon_{L_i} \epsilon_{E_j}$, so that each 
matter field carries its own suppression factor.  This arises in 
many models of flavor, for example in classes of models with flavor 
symmetries or strong dynamics.  The important point is the similarity 
between the forms of $(y_f)_{ij}$ and $(a_f)_{ij}$.  This comes from 
the fact that matter and Higgs fields appear identically in the two 
classes of operators in Eq.~(\ref{eq:Yukawa-sector}).

The gaugino masses, which arise from ${\cal O}_{\lambda_A}$ as 
$M_A = -2 \eta_A g_A^2 F_X/M$, set the scale for the superparticle 
masses.  Assuming that the mediation mechanism produces unsuppressed 
${\cal O}_{\zeta_f}$, i.e. $\tilde{\zeta} \approx \eta_A$, we then obtain 
$(a_f)_{ij} \approx {\cal E}^f_{ij} M_A/g_A^2$ at the scale $M_F$.%
\footnote{Note that in our notation, the grand unified relations for 
 the gaugino masses correspond to $\eta_1 = \eta_2 = \eta_3$.}
Taking into account Eq.~(\ref{eq:y_f-E}) and SSM renormalization 
group evolution below $M$, we can write the flavor nonuniversal 
part of the low energy scalar trilinear interactions as
\begin{equation}
  (a_u)_{ij} \approx (y_u)_{ij} \frac{a_{\rm C}}{\tilde{y}},
\qquad
  (a_d)_{ij} \approx (y_d)_{ij} \frac{a_{\rm C}}{\tilde{y}},
\qquad
  (a_e)_{ij} \approx (y_e)_{ij} \frac{a_{\rm N}}{\tilde{y}},
\label{eq:a_f-y_f}
\end{equation}
where $a_{\rm C}, a_{\rm N} \approx O(M_A)$ are characteristic mass 
scales for these interactions associated with colored and non-colored 
sfermions, and we expect $a_{\rm C} \simgt a_{\rm N}$ due to the 
structure of the SSM renormalization group equations.  We note 
again that $O(1)$ coefficients are omitted for each element of 
Eq.~(\ref{eq:a_f-y_f}), so that $(a_f)_{ij}$ is not proportional 
to $(y_f)_{ij}$ as a matrix.

\subsection{The superpotential flavor problem}
\label{subsec:constraints}

The flavor nonuniversality at the level of Eq.~(\ref{eq:a_f-y_f}) 
can be problematic.  Flavor and $CP$ violation can in general be 
quantified by mass insertion parameters, which are obtained by dividing 
the off-diagonal entry of the sfermion mass-squared matrix by the average 
diagonal entry in the super-CKM basis~\cite{Dugan:1984qf,Gabbiani:1996hi}. 
The mass insertion parameters obtained from Eq.~(\ref{eq:a_f-y_f}) are
\begin{equation}
  (\delta^u_{LR})_{ij} 
    \approx \frac{a_{\rm C}}{\tilde{y}\, m_{\rm C}^2} (M_u)_{ij},
\qquad
  (\delta^d_{LR})_{ij} 
    \approx \frac{a_{\rm C}}{\tilde{y}\, m_{\rm C}^2} (M_d)_{ij},
\qquad
  (\delta^e_{LR})_{ij} 
    \approx \frac{a_{\rm N}}{\tilde{y}\, m_{\rm N}^2} (M_e)_{ij},
\label{eq:delta-LR-ij}
\end{equation}
for $i \neq j$, where $m_{\rm C}$ and $m_{\rm N}$ are characteristic 
masses for colored and non-colored superparticles, and we expect 
$m_{\rm C} \simgt m_{\rm N}$.  Here, $(M_u)_{ij} = (y_u)_{ij} \langle 
H_u \rangle$, $(M_d)_{ij} = (y_d)_{ij} \langle H_d \rangle$ and 
$(M_e)_{ij} = (y_e)_{ij} \langle H_d \rangle$ are the quark and lepton 
mass matrices in the original (not super-CKM) basis.  The diagonal 
elements, $(\delta^f_{LR})_{ii}$, receive additional terms coming from 
the flavor universal contribution to $(a_f)_{ij}$ from renormalization 
group evolution below $M$ and the contribution to the sfermion 
left-right masses proportional to $|\mu|$.  However, assuming 
there is no intrinsic $CP$ violation associated with supersymmetry 
breaking, these terms do not contribute to the imaginary parts of 
$(\delta^f_{LR})_{ii}$, which are relevant in the discussion below. 
We thus find
\begin{equation}
  {\rm Im}(\delta^u_{LR})_{ii} \approx 
    \frac{a_{\rm C} \sin\varphi_u}{\tilde{y}\, m_{\rm C}^2} (M_u)_{ii},
\qquad
  {\rm Im}(\delta^d_{LR})_{ii} \approx 
    \frac{a_{\rm C} \sin\varphi_d}{\tilde{y}\, m_{\rm C}^2} (M_d)_{ii},
\qquad
  {\rm Im}(\delta^e_{LR})_{ii} \approx 
    \frac{a_{\rm N} \sin\varphi_e}{\tilde{y}\, m_{\rm N}^2} (M_e)_{ii},
\label{eq:delta-LR-ii}
\end{equation}
where $\varphi_f$ are the phases of the contributions to 
$(\delta^f_{LR})_{ii}$ from the flavor nonuniversal part of 
$(a_f)_{ij}$.  Note that $\varphi_f \approx O(1)$ is expected 
even if supersymmetry breaking does not introduce new $CP$ violating 
phases because these complex phases arise generically from the 
Yukawa couplings when going into the super-CKM basis.%
\footnote{This may be avoided in certain models, e.g., 
 models with hermitian Yukawa and scalar trilinear interaction 
 matrices~\cite{Kuchimanchi:1995rp} and those with spontaneous 
 $CP$ violation~\cite{Nir:1996am}.}

The theoretical estimate of Eqs.~(\ref{eq:delta-LR-ij}, \ref{eq:delta-LR-ii}) 
can be compared with experimental constraints from low energy observables. 
The bound on the $\mu \rightarrow e\gamma$ process~\cite{Brooks:1999pu} 
gives
\begin{equation}
  \frac{1}{\sqrt{2}} \sqrt{|(\delta^e_{LR})_{12}|^2 
    + |(\delta^e_{LR})_{21}|^2} \simlt 4 \times 10^{-6} 
    \left( \frac{m_{\rm N}}{200~{\rm GeV}} \right).
\label{eq:mu-e-gamma}
\end{equation}
The $\mu \rightarrow e$ conversion and $\mu \rightarrow eee$ processes 
also give comparable bounds.  The limits on the electric dipole moments 
(EDMs) of the electron~\cite{Regan:2002ta}, neutron~\cite{Baker:2006ts} 
and mercury atom~\cite{Romalis:2000mg} lead to
\begin{equation}
  |{\rm Im}(\delta^e_{LR})_{11}| \simlt 2 \times 10^{-7} 
    \left( \frac{m_{\rm N}}{200~{\rm GeV}} \right),
\label{eq:e-EDM}
\end{equation}
\begin{equation}
  |{\rm Im}(\delta^u_{LR})_{11}| \simlt 2 \times 10^{-6} 
    \left( \frac{m_{\rm C}}{600~{\rm GeV}} \right),
\qquad
  |{\rm Im}(\delta^d_{LR})_{11}| \simlt 1 \times 10^{-6} 
    \left( \frac{m_{\rm C}}{600~{\rm GeV}} \right),
\label{eq:n-EDM}
\end{equation}
\begin{equation}
  |{\rm Im}(\delta^u_{LR})_{11}| \simlt 4 \times 10^{-7} 
    \left( \frac{m_{\rm C}}{600~{\rm GeV}} \right),
\qquad
  |{\rm Im}(\delta^d_{LR})_{11}| \simlt 4 \times 10^{-7} 
    \left( \frac{m_{\rm C}}{600~{\rm GeV}} \right),
\label{eq:Hg-EDM}
\end{equation}
respectively.  Here, the bounds of 
Eqs.~(\ref{eq:mu-e-gamma}~--~\ref{eq:Hg-EDM}) are obtained 
conservatively by scanning the ratios of the superparticle masses 
in a reasonable range (see e.g.~\cite{Gabbiani:1996hi,Abel:2001vy}).%
\footnote{Here we consider the ranges $m_{\tilde{g}}^2/m_{\tilde{q}}^2 
 \simlt 2$ and $m_\chi^2/m_{\tilde{l}}^2 \simlt 3$, where $m_{\tilde{g}}$, 
 $m_{\tilde{q}}$, $m_\chi$, and $m_{\tilde{l}}$ are the gluino, squark, 
 weak gaugino, and slepton average masses.  These ranges are motivated 
 by renormalization group considerations with $M_F$ well above the 
 TeV scale, e.g. $M_F \simgt 10^{10}~{\rm GeV}$.  The case of smaller 
 $M_F$ will be discussed in section~\ref{subsec:general-sol}.}
The bounds from neutron and mercury EDMs are subject to large 
theoretical uncertainties~\cite{Falk:1999tm}, and we have used 
conservative estimates.  The constraints from $(\epsilon'/\epsilon)_K$ 
and $b \rightarrow s \gamma$ also lead to bounds on 
$|{\rm Im}(\delta^d_{LR})_{12,\,21}|$ and $|(\delta^d_{LR})_{23,\,32}|$, 
but they are not as strong as the bounds above when the left-right 
mass insertion parameters scale naively with the quark masses.

We can obtain the bounds on the superparticle masses using 
Eqs.~(\ref{eq:delta-LR-ij}, \ref{eq:delta-LR-ii}), with the 
approximation $a_{\rm C} \approx m_{\rm C}$ and $a_{\rm N} \approx 
m_{\rm N}$, which is sufficient for the level of analysis here. 
Taking $(M_u)_{11} \simeq m_u \simeq 2~{\rm MeV}$ and $(M_d)_{11} \simeq 
m_d \simeq 4~{\rm MeV}$, the neutron EDM bound of Eq.~(\ref{eq:n-EDM}) 
leads to the following bound on $m_C$:
\begin{equation}
  m_{\rm C} \simgt {\rm max}\left\{ 
    800~{\rm GeV} \left( \frac{\sin\varphi_u}{\tilde{y}} \right)^{1/2}, 
  \,\,\, 
    1.5~{\rm TeV} \left( \frac{\sin\varphi_d}{\tilde{y}} \right)^{1/2} 
  \right\},
\label{eq:mC-bounds-1}
\end{equation}
whereas the mercury EDM bound, Eq.~(\ref{eq:Hg-EDM}), gives
\begin{equation}
  m_{\rm C} \simgt {\rm max}\left\{ 
    1.3~{\rm TeV} \left( \frac{\sin\varphi_u}{\tilde{y}} \right)^{1/2}, 
  \,\,\, 
    1.9~{\rm TeV} \left( \frac{\sin\varphi_d}{\tilde{y}} \right)^{1/2} 
  \right\}.
\label{eq:mC-bounds-2}
\end{equation}
The bound on $m_N$ depends on the assumption on the charged lepton 
mass matrix.  If we conservatively take $(M_e)_{11} \simeq m_e 
\simeq 0.5~{\rm MeV}$ and $(M_e)_{12} \simeq (M_e)_{21} \simeq 
(m_e\, m_\mu)^{1/2} \simeq 7~{\rm MeV}$, we obtain
\begin{equation}
  m_{\rm N} \simgt {\rm max}\left\{ 
    600~{\rm GeV} \frac{1}{\tilde{y}^{1/2}}, 
  \,\,\, 
    700~{\rm GeV} \left( \frac{\sin\varphi_e}{\tilde{y}} \right)^{1/2} 
  \right\}.
\label{eq:mN-bounds-1}
\end{equation}
On the other hand, if the large neutrino mixing angle $\theta_{12}$ 
receives a significant contribution from the charged lepton Yukawa 
matrix, we expect $(M_e)_{12} \simeq m_\mu \tan\theta_{12} \simeq 
70~{\rm MeV}$, giving a much stronger bound
\begin{equation}
  m_{\rm N} \simgt {\rm max}\left\{ 
    1.9~{\rm TeV} \frac{1}{\tilde{y}^{1/2}}, 
  \,\,\, 
    700~{\rm GeV} \left( \frac{\sin\varphi_e}{\tilde{y}} \right)^{1/2} 
  \right\}.
\label{eq:mN-bounds-2}
\end{equation}
In fact, this latter bound is expected to apply in the case where 
${\cal E}^e_{ij}$ factorizes: ${\cal E}^e_{ij} = \epsilon_{L_i} 
\epsilon_{E_j}$, since then the large 1-2 neutrino mixing angle 
generically implies $\epsilon_{L_1} \approx \epsilon_{L_2}$, leading 
to a large 1-2 element of the charged lepton mass matrix of $O(m_\mu)$.

In addition to the uncertainties already described, the bounds on 
$m_{\rm C,N}$ derived above are subject to uncertainties coming from 
$O(1)$ coefficients in front of Eq.~(\ref{eq:a_f-y_f}).  Only the square 
root of these coefficients, however, appear in the bounds.  For example, 
if we take the magnitude of these coefficients to be between $0.5$ and 
$2$, the bounds receive unknown coefficients of $O(0.7$\,--\,$1.4)$, which 
do not significantly affect the results.  These bounds also scale with 
the square root of the natural size of the scalar trilinear interactions 
at $M$, $(\tilde{\zeta}/\eta_A)^{1/2}$, which we have set unity. 
In addition, the bounds scale with the experimental limits on the 
$\mu \rightarrow e \gamma$ branching ratio, ${\rm Br}(\mu \rightarrow 
e \gamma)$, and the electron, neutron and mercury EDMs, $d_e$, $d_n$ 
and $d_{\rm Hg}$, as
\begin{equation}
  \left( \frac{{\rm Br}(\mu \rightarrow e \gamma)}{1.2 \times 10^{-11}} 
    \right)^{-1/4},
\label{eq:scaling-1}
\end{equation}
\begin{equation}
  \left( \frac{d_e}{1.6 \times 10^{-27}\, e\,{\rm cm}} \right)^{-1/2},
\qquad
  \left( \frac{d_n}{2.9 \times 10^{-26}\, e\,{\rm cm}} \right)^{-1/2},
\qquad
  \left( \frac{d_{\rm Hg}}{2.1 \times 10^{-28}\, e\,{\rm cm}} \right)^{-1/2}.
\label{eq:scaling-2}
\end{equation}
Therefore, if future experiments such as ones in 
Refs.~[\ref{MEG:X}~--~\ref{Semertzidis:2003iq:X}] improve 
the upper bounds on these (and other) quantities, the lower 
bounds on $m_{\rm C,N}$ increase accordingly.

The bounds of Eqs.~(\ref{eq:mC-bounds-1}~--~\ref{eq:mN-bounds-2}) place 
lower limits on the superparticle masses, yielding a tension with the 
observability of supersymmetry at the LHC. In fact, the conservative 
bound of Eq.~(\ref{eq:mN-bounds-1}) already gives strong constraints 
on the superparticle spectrum for $\tilde{y} \approx 1$.  In particular, 
in the case that colored and non-colored superparticles do not have 
a strong mass hierarchy at $M \gg {\rm TeV}$, we expect that $m_{\rm C} 
\approx (2$\,--\,$4) m_{\rm N}$ at low energies.  This pushes up the 
masses of colored superparticles beyond $1~{\rm TeV}$, and, in many 
cases, beyond the reach of the LHC.  The constraints are even stronger 
if the large neutrino mixing angle $\theta_{12}$ receives a sizable 
contribution from the charged lepton Yukawa matrix, as in 
Eq.~(\ref{eq:mN-bounds-2}).  We call this generic tension between 
low energy flavor and $CP$ violation and the observability of 
supersymmetry at the LHC {\it the superpotential flavor problem}, 
since it is caused by the superpotential operators ${\cal O}_{\zeta_f}$. 
An important point, again, is that the problem is relatively model 
independent because the flavor structure of ${\cal O}_{\zeta_f}$ 
is expected to be correlated with that of ${\cal O}_{\lambda_f}$ 
in wide classes of flavor theories.

\subsection{More general problem with left-right sfermion propagation}
\label{subsec:general-LR}

The superpotential flavor problem provides a strong, model-independent 
tension between weak scale supersymmetry and low energy flavor and 
$CP$ violating observables.  This is, however, only one aspect of 
a more general problem associated with left-right propagation of 
the sfermions in flavor and $CP$ violating amplitudes.

\begin{figure}[t]
\begin{center} 
\begin{picture}(480,90)(0,48)
  \Text(60,61)[t]{\large (a)}
  \Line(8,90)(30,90)   \Text(9,86)[tl]{$f_L$}
  \Line(90,90)(112,90) \Text(111,86)[tr]{$f_R$}
  \Line(30,90)(90,90) \Photon(30,90)(90,90){3}{7}
  \Text(60,81)[t]{$\lambda$}
  \DashCArc(60,90)(30,0,180){3} \Text(60,115)[t]{$\tilde{f}$}
  \Line(42.0,113.0)(48.0,119.0) \Line(42.0,119.0)(48.0,113.0)
  \Text(42.0,121.2)[br]{$\delta^f_{LL}$}
  \Line(72.0,113.0)(78.0,119.0) \Line(72.0,119.0)(78.0,113.0)
  \Text(78.0,121.2)[bl]{$\delta^f_{LR}$}
  \Text(180,61)[t]{\large (b)}
  \Line(128,90)(150,90) \Text(129,86)[tl]{$f_L$}
  \Line(210,90)(232,90) \Text(231,86)[tr]{$f_R$}
  \Line(150,90)(210,90) \Photon(150,90)(210,90){3}{7}
  \Text(180,81)[t]{$\lambda$}
  \DashCArc(180,90)(30,0,180){3} \Text(180,115)[t]{$\tilde{f}$}
  \Line(162.0,113.0)(168.0,119.0) \Line(162.0,119.0)(168.0,113.0)
  \Text(162.0,121.2)[br]{$\delta^f_{LR}$}
  \Line(192.0,113.0)(198.0,119.0) \Line(192.0,119.0)(198.0,113.0)
  \Text(198.0,121.2)[bl]{$\delta^f_{RR}$}
  \Text(300,61)[t]{\large (c)}
  \Line(248,90)(270,90) \Text(249,86)[tl]{$f_L$}
  \Line(330,90)(352,90) \Text(351,86)[tr]{$f_R$}
  \Line(270,90)(330,90) \Photon(270,90)(330,90){3}{7}
  \Text(300,81)[t]{$\lambda$}
  \DashCArc(300,90)(30,0,180){3} \Text(300,115)[t]{$\tilde{f}$}
  \Line(275.8,108.2)(281.8,114.2) \Line(275.8,114.2)(281.8,108.2)
  \Text(275.5,115.5)[br]{$\delta^f_{LL}$}
  \Line(297.0,117.0)(303.0,123.0) \Line(297.0,123.0)(303.0,117.0)
  \Text(300.0,129.1)[b]{$\delta^f_{LR}$}
  \Line(318.2,108.2)(324.2,114.2) \Line(318.2,114.2)(324.2,108.2)
  \Text(326.5,115.5)[bl]{$\delta^f_{RR}$}
  \Text(420,61)[t]{\large (d)}
  \Line(368,90)(390,90) \Text(369,86)[tl]{$f_L$}
  \Line(450,90)(472,90) \Text(471,86)[tr]{$f_R$}
  \Line(390,90)(450,90) \Photon(390,90)(450,90){3}{7}
  \Text(420,81)[t]{$\lambda$}
  \DashCArc(420,90)(30,0,180){3} \Text(420,115)[t]{$\tilde{f}$}
  \Line(395.8,108.2)(401.8,114.2) \Line(395.8,114.2)(401.8,108.2)
  \Text(395.5,115.5)[br]{$\delta^f_{LR}$}
  \Line(417.0,117.0)(423.0,123.0) \Line(417.0,123.0)(423.0,117.0)
  \Text(420.0,129.1)[b]{$\delta^f_{RL}$}
  \Line(438.2,108.2)(444.2,114.2) \Line(438.2,114.2)(444.2,108.2)
  \Text(446.5,115.5)[bl]{$\delta^f_{LR}$}
\end{picture}
\caption{Multiple mass insertion diagrams that lead to dangerous flavor 
 and $CP$ violating contributions.  Here, $f_{L,R}$, $\tilde{f}$ and 
 $\lambda$ represent fermions, scalars and gauginos, respectively.}
\label{fig:multiple}
\end{center}
\end{figure}
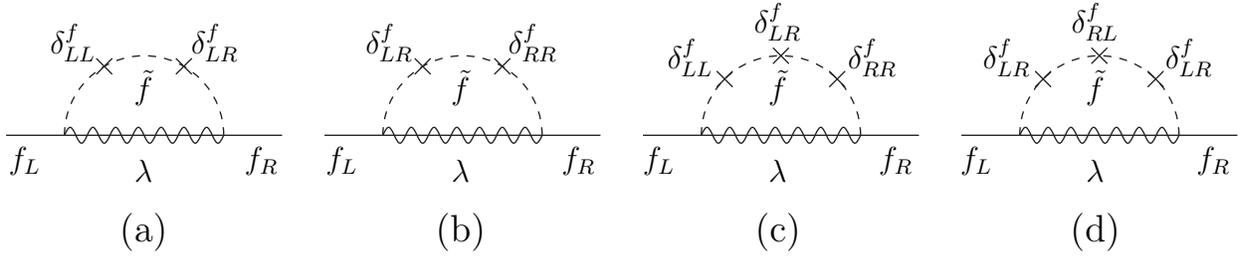

Suppose that the superpotential flavor problem is somehow solved, 
i.e. the operators ${\cal O}_{\zeta_f}$ are strongly suppressed.  There 
will still be the contributions to lepton flavor violation and EDMs 
associated with left-right propagation of sfermions.  First of all, 
there are flavor nonuniversal scalar trilinear interactions generated 
by ${\cal O}_{Z_\Phi,\eta_\Phi}$, yielding $(\delta^f_{LR})_{ij}$ 
($i \neq j$) and ${\rm Im}(\delta^f_{LR})_{ii}$.  These contributions 
must be sufficiently suppressed.  Moreover, even if they are small, 
lepton flavor violation and EDMs are induced by diagrams that 
use multiple mass insertion parameters $(\delta^f_{LR})_{ij}$, 
$(\delta^f_{LL})_{ij}$ and $(\delta^f_{RR})_{ij}$ (see 
Fig.~\ref{fig:multiple})~\cite{Feng:2001sq}, instead of a single 
insertion of $(\delta^f_{LR})_{ij}$.  Since the diagrams depend on 
parameters $(\delta^f_{LL})_{ij}$ and $(\delta^f_{RR})_{ij}$, whose 
correlations with the Yukawa couplings are model dependent, the 
tension caused by these diagrams is not as model independent as 
the superpotential flavor problem.  Nevertheless, this provides 
strong constraints on supersymmetric models in which the structure 
of the supersymmetry breaking parameters is correlated with that 
of the Yukawa couplings.

One might naively think that because of the use of multiple mass 
insertion parameters, the diagrams of Fig.~\ref{fig:multiple} are 
much smaller than those using a single $(\delta^f_{LR})_{ij}$. 
This is, however, not always the case for the following reasons:
\begin{itemize}
\item
The left-right mass insertions used can be flavor universal, 
${\rm Re}(\delta^f_{LR})_{ii}$, since the necessary flavor/$CP$ 
violation can come from insertions of $(\delta^f_{LL})_{ij}$ and/or 
$(\delta^f_{RR})_{ij}$.  This may enhance the contributions from 
multiple mass insertion diagrams relative to single insertion 
ones, especially for $f=d,e$, since the flavor universal part of 
$(\delta^{d,e}_{LR})_{ij}$ is enhanced by $\tan\beta \equiv \langle 
H_u \rangle/\langle H_d \rangle$.  (For $f=e$, it is also enhanced 
by $\mu/m_{\rm N}$, which is typically of $O(m_{\rm C}/m_{\rm N})$.)
\item
The sfermions propagating between two mass insertions can be from 
a heavier generation.  For diagrams with triple mass insertions, for 
example, the states propagating between mass insertions can be third 
generation states, minimizing extra suppressions arising from use 
of more mass insertion parameters.
\end{itemize}
In fact, these two ingredients can make the contributions from the 
diagrams of Fig.~\ref{fig:multiple} comparable or even larger than those 
from the diagrams with a single insertion of $(\delta^f_{LR})_{ij}$ 
of Eqs.~(\ref{eq:delta-LR-ij}, \ref{eq:delta-LR-ii}).

To illustrate this point, suppose that ${\cal E}^f_{ij}$ in 
Eq.~(\ref{eq:y_f-E}) factorize, ${\cal E}^u_{ij} = \epsilon_{Q_i} 
\epsilon_{U_j}$, ${\cal E}^d_{ij} = \epsilon_{Q_i} \epsilon_{D_j}$ and 
${\cal E}^e_{ij} = \epsilon_{L_i} \epsilon_{E_j}$, giving the Yukawa 
couplings $(y_u)_{ij} \approx \tilde{y}\, \epsilon_{Q_i} \epsilon_{U_j}$, 
$(y_d)_{ij} \approx \tilde{y}\, \epsilon_{Q_i} \epsilon_{D_j}$ and 
$(y_e)_{ij} \approx \tilde{y}\, \epsilon_{L_i} \epsilon_{E_j}$.  Suppose 
also that the flavor nonuniversal part of the sfermion squared masses 
scale naively with the $\epsilon$ factors:
\begin{equation}
  (m_\Phi^2)_{ij} \approx 
    \epsilon_{\Phi_i} \epsilon_{\Phi_j} m_{\rm S}^2,
\label{eq:m2-naive}
\end{equation}
where $m_{\rm S} \approx \eta_A F_X/M$ is the scale of supersymmetry 
breaking parameters at $M$, which we assume to be the same for colored 
and non-colored superparticles.  The mass insertion parameters generated 
by Eq.~(\ref{eq:m2-naive}) are then
\begin{equation}
  (\delta^u_{LL})_{ij} \approx (\delta^d_{LL})_{ij} \approx 
    \epsilon_{Q_i} \epsilon_{Q_j} \frac{m_{\rm S}^2}{m_{\rm C}^2},
\qquad
  (\delta^u_{RR})_{ij} \approx 
    \epsilon_{U_i} \epsilon_{U_j} \frac{m_{\rm S}^2}{m_{\rm C}^2},
\qquad
  (\delta^d_{RR})_{ij} \approx 
    \epsilon_{D_i} \epsilon_{D_j} \frac{m_{\rm S}^2}{m_{\rm C}^2},
\label{eq:LL-RR-squarks}
\end{equation}
\begin{equation}
  (\delta^e_{LL})_{ij} \approx (\delta^\nu_{LL})_{ij} \approx 
    \epsilon_{L_i} \epsilon_{L_j} \frac{m_{\rm S}^2}{m_{\rm N}^2},
\qquad
  (\delta^e_{RR})_{ij} \approx 
    \epsilon_{E_i} \epsilon_{E_j} \frac{m_{\rm S}^2}{m_{\rm N}^2},
\label{eq:LL-RR-sleptons}
\end{equation}
where $i \neq j$, and we have included the mass insertion parameters 
for the sneutrinos.  On the other hand, the dominant contribution to 
the flavor universal part of the left-right mass insertion parameters 
are given by
\begin{equation}
  {\rm Re}(\delta^u_{LR})_{ii} 
    \approx \frac{1}{m_{\rm C}} (M_u)_{ii},
\qquad
  {\rm Re}(\delta^d_{LR})_{ii} 
    \approx \frac{\mu \tan\beta}{m_{\rm C}^2} (M_d)_{ii},
\qquad
  {\rm Re}(\delta^e_{LR})_{ii} 
    \approx \frac{\mu \tan\beta}{m_{\rm N}^2} (M_e)_{ii},
\label{eq:delta-Re_LR-ii}
\end{equation}
where we have taken $\mu \approx m_{\rm C} \simgt m_{\rm N}$ and 
$\tan\beta \simgt 1$, and assumed that flavor universal scalar 
trilinear interactions $(a_{u,d})_{ii} \approx (y_{u,d})_{ii} 
m_{\rm C}$ and $(a_e)_{ii} \approx (y_e)_{ii} m_{\rm N}$ are 
generated by renormalization group evolution.  (The expression 
of Eq.~(\ref{eq:delta-Re_LR-ii}) also applies to the case 
where ${\cal E}^f_{ij}$ do not factorize.)

Consider, for example, the diagram of Fig.~\ref{fig:multiple}(c) with 
$f = u$.  This leads to the contribution to the up quark EDM that 
scales with
\begin{equation}
  (\delta^u_{LL})_{13} (\delta^u_{LR})_{33} (\delta^u_{RR})_{31} 
    \approx \frac{(M_u)_{11}}{\tilde{y}\, m_{\rm C}} 
      \left( \frac{m_{\rm S}}{m_{\rm C}} \right)^4 
      \frac{(y_u)_{33}^2}{\tilde{y}},
\label{eq:ex-triple}
\end{equation}
which can be comparable to the dangerous contribution that scales 
with $(\delta^u_{LR})_{11} \approx (M_u)_{11}/\tilde{y}\, m_{\rm C}$ 
of Eq.~(\ref{eq:delta-LR-ij}) with $a_{\rm C} \approx m_{\rm C}$. 
The diagram of Fig.~\ref{fig:multiple}(a) with $f = e$ gives 
a contribution to the $\mu \rightarrow e \gamma$ process that 
scales with
\begin{equation}
  (\delta^e_{LL})_{12} (\delta^e_{LR})_{22} 
    \approx \frac{(M_e)_{12}}{\tilde{y}\, m_{\rm N}}\,\, 
      \frac{m_{\rm S}^2\, \mu}{m_{\rm N}^3}\, 
      \tilde{y}\, \epsilon_{L_2}^2 \tan\beta,
\label{eq:ex-double}
\end{equation}
which can also be dangerous because it could be comparable to the 
contribution from $(\delta^e_{LR})_{12} \approx (M_e)_{12}/\tilde{y}\, 
m_{\rm N}$ of Eq.~(\ref{eq:delta-LR-ij}) with $a_{\rm N} \approx 
m_{\rm N}$, especially for large $\tan\beta$.  These examples show 
that the multiple mass insertion diagrams may lead to flavor and $CP$ 
violation at a dangerous level even in the absence of flavor and $CP$ 
violating $(\delta^f_{LR})_{ij}$.

In practice, the constraints from multiple mass insertion diagrams 
can be taken into account by considering the effective left-right mass 
insertion parameters
\begin{eqnarray}
  (\delta^f_{LR,{\rm eff}})_{ij} &\equiv& 
    {\rm max}\Bigl\{ c_d (\delta^f_{LL})_{ik} (\delta^f_{LR})_{kj},\,\, 
    c_d (\delta^f_{LR})_{ik} (\delta^f_{RR})_{kj},\,\,
\nonumber\\
  && \qquad
    c_t (\delta^f_{LL})_{ik} (\delta^f_{LR})_{kk} (\delta^f_{RR})_{kj},\,\,
    c_t (\delta^f_{LR})_{ik} (\delta^f_{RL})_{kk} (\delta^f_{LR})_{kj} 
  \Bigr\},
\label{eq:delta-LR_eff}
\end{eqnarray}
and requiring that $(\delta^f_{LR,{\rm eff}})_{ij}$ satisfy the 
bounds of Eqs.~(\ref{eq:mu-e-gamma}~--~\ref{eq:Hg-EDM}) with 
$(\delta^f_{LR})_{ij}$ replaced by $(\delta^f_{LR,{\rm eff}})_{ij}$. 
Here, $c_d \simeq (0.5$\,--\,$0.8)$ and $c_t \simeq (0.3$\,--\,$0.6)$ 
are numerical coefficients arising from the difference of momentum 
integral functions with various numbers of insertions.  Once 
$(\delta^f_{LL})_{ij}$ and $(\delta^f_{RR})_{ij}$ are given, 
these constraints can be checked.

\section{Approaches to the Problem}
\label{sec:solutions}

In order to have a framework for weak scale supersymmetry in which the 
LHC can provide additional insight into the origin of the observed flavor 
structure, the supersymmetric left-right flavor problem must somehow 
be addressed.  The bounds associated with left-left and right-right 
sfermion propagation must also be avoided, although they are, in general, 
less stringent.  To address the issue of whether there are theories 
that naturally satisfy all these constraints, we start by identifying 
classes of theories that do not have the superpotential flavor problem, 
a robust part of the supersymmetric left-right flavor problem.  While 
it is not automatic that these theories will be safe from low energy 
constraints or even solve the supersymmetric left-right flavor problem, 
they provide frameworks with which to build more detailed theories 
that can avoid all low energy constraints.  The remaining constraints 
will be discussed in the next section.

\subsection{General considerations}
\label{subsec:general-sol}

There are essentially two different directions to address the 
superpotential flavor problem.  One is to assume that the operators 
${\cal O}_{\zeta_f}$ exist with their natural size, but the bounds 
are somehow avoided.  Barring accidental cancellations in the amplitudes 
for low energy flavor and $CP$ violating effects, this includes the 
following possibilities:
\begin{enumerate}
\item[(i)]
The bounds are given by Eq.~(\ref{eq:mN-bounds-1}), and the superparticles 
are not too much heavier.  If the superparticle masses satisfy 
$m_{\rm C} \approx (2$\,--\,$4) m_{\rm N}$, the viable parameter 
region is somewhat squeezed.  The constraints are slightly relaxed 
if we allow the masses of colored and non-colored superparticles to 
be of similar size, $m_{\rm C} \sim m_{\rm N}$.  Avoiding the bound 
of Eq.~(\ref{eq:mN-bounds-2}), however, is still not easy.
\item[(ii)]
The intrinsic size of the Yukawa couplings is large, $\tilde{y} 
\gg O(1)$.  In this case the bounds on $m_C$ and $m_N$ are not 
significant, especially in the case where the large neutrino mixing 
angle $\theta_{12}$ arises only from the neutrino mass matrix, 
Eqs.~(\ref{eq:mC-bounds-1}~--~\ref{eq:mN-bounds-1}).  If the bound 
is given by Eq.~(\ref{eq:mN-bounds-2}), the lower bound on $m_{\rm N}$ 
can be relaxed to about $500~{\rm GeV}$ by taking the largest possible 
value of $\tilde{y} \approx 4\pi$.  This constraint, however, is 
still significant.
\item[(iii)]
The gauginos are significantly heavier than the sfermions.  In this 
case the bounds of Eqs.~(\ref{eq:mC-bounds-1}, \ref{eq:mC-bounds-2}) 
and Eqs.~(\ref{eq:mN-bounds-1}, \ref{eq:mN-bounds-2}) are relaxed 
approximately by a factor of $(m_{\tilde{g}}/m_{\tilde{q}})/\sqrt{2}$ 
and $(m_\chi/m_{\tilde{l}})/\sqrt{3}$, respectively, with $m_{\rm C,N}$ 
now interpreted as the masses of the sfermions.  This situation can 
occur if $M_F$ is close to the TeV scale, and the masses of the sfermions 
at $M_F$ are suppressed by the dynamics generating the Yukawa hierarchy, 
as in flavor models of Ref.~\cite{Nelson:2000sn}.%
\footnote{$M_F$ needs to be low to prevent the scalar masses from 
 becoming comparable to the gaugino masses through SSM renormalization 
 group evolution.}
Note that we only need sfermions to be accessible at the LHC to probe 
the origin of the flavor structure.
\end{enumerate}
These possibilities are certainly viable, especially given uncertainties 
in our estimates.  The tension between flavor constraints and LHC 
observability, however, still exists.  If one of these possibilities 
is realized, and the superparticles are within the LHC reach, we expect 
that $\mu \rightarrow e$ processes and/or atomic and nuclear EDMs will 
be discovered in the near future, for example in the experiments 
of Refs.~[\ref{MEG:X}~--~\ref{Semertzidis:2003iq:X}], which expect 
to improve present bounds by several orders of magnitude.

The other direction to address the superpotential flavor problem 
is to consider that the operators ${\cal O}_{\zeta_f}$ are somehow 
suppressed.  This includes the following possibilities:
\begin{enumerate}
\item[(iv)]
The coefficients of the operators ${\cal O}_{\zeta_f}$ (or at least 
those of the 1-2 and 2-1 elements of ${\cal O}_{\zeta_e}$) are 
accidentally suppressed.  The required amount of suppression is not 
strong if we adopt Eq.~(\ref{eq:mN-bounds-1}).  However, if we instead 
use Eq.~(\ref{eq:mN-bounds-2}), we need to have $\tilde{\zeta}/\eta_A 
\simlt 0.07\, \tilde{y}\, (m_{\rm N}/500~{\rm GeV})^2$, which provides 
a strong bound on $\tilde{\zeta}/\eta_A$ for $\tilde{y} \approx 1$.
\item[(v)]
The scalar trilinear interactions are exactly proportional to the Yukawa 
matrices, $(a_f)_{ij} \propto (y_f)_{ij}$, leading to vanishing flavor 
and $CP$ violating mass insertion parameters.  This may be achieved, 
for example, if $(a_f)_{ij}$ and $(y_f)_{ij}$ arise from a single 
operator through the lowest and highest components VEVs of $X$, 
$\langle X \rangle = X_0 + \theta^2 F_X$, with ${\rm arg}(F_X/X_0) 
\approx {\rm arg}(M_A)$.  The large top quark mass, however, requires 
that $X_0$ is close to the cutoff scale $M_*$, and the problem may 
be regenerated by higher order terms in $X_0/M_*$.
\item[(vi)]
The operators ${\cal O}_{\zeta_f}$ are suppressed by some mechanism, 
$\tilde{\zeta} \ll \eta_A$.  This mechanism may or may not operate in 
the regime where effective field theory is valid.
\end{enumerate}
Note that (iv), (v) and (vi) above can also be combined with (i), 
(ii) and (iii) described before.  For example, we can consider a 
setup where ${\cal O}_{\zeta_f}$ are suppressed by some mechanism, 
(vi), {\it and} the natural size of the Yukawa couplings, $\tilde{y}$, 
is large, (ii).

In the rest of this paper, we focus on the last possibility, (vi), 
and see how much the situation will be improved.  As we discussed, we 
still need to address the more general left-right flavor problem and 
the constraints from left-left and right-right sfermion propagation, 
which we defer to the next section.  Here we present simple classes 
of theories in which ${\cal O}_{\zeta_f}$ are naturally suppressed. 
In fact, this provides a platform for the analysis of the more model 
dependent part of the flavor problem.

There are several possible ways that the operators ${\cal O}_{\zeta_f}$ 
can be suppressed.  In fact, they may simply be absent at a scale 
where the SSM arises as an effective field theory, as a result of 
the dynamics of some more fundamental theory.  This is not unreasonable 
because ${\cal O}_{\zeta_f}$ are the only superpotential operators 
associated with supersymmetry breaking, so if supersymmetry breaking 
is mediated to the SSM sector by loop processes then these operators 
may be absent.  In the remainder of this section we discuss three 
simple classes of theories for suppressing ${\cal O}_{\zeta_f}$. 
We classify them according to the pattern of the supersymmetry 
breaking parameters obtained at $M_F$.  The simple suppression of 
${\cal O}_{\zeta_f}$ described above can effectively be classified 
into the first class, because it leads to the same pattern of the 
supersymmetry breaking masses at $M_F$.  There are clearly many 
models within each class, and we explicitly discuss some of them.

\subsection{Framework I --- Higgsphobic supersymmetry breaking}
\label{subsec:framework-1}

We here present the first class of theories in which ${\cal O}_{\zeta_f}$ 
are naturally suppressed.  A unique feature of the operators 
${\cal O}_{\zeta_f}$ is that among the operators relevant to flavor 
violation, Eqs.~(\ref{eq:matter-sector-1}~--~\ref{eq:Yukawa-sector}), 
these are the only operators that contain both Higgs fields, 
$H$, and the supersymmetry breaking field, $X$.  Therefore, if 
the theory does not allow direct coupling between $H$ and $X$, 
a framework which we call {\it Higgsphobic supersymmetry breaking}, 
then ${\cal O}_{\zeta_f}$ are forbidden.  The other operators 
${\cal O}_{Z_\Phi,\kappa_\Phi,\eta_\Phi,\lambda_f}$ can exist 
as long as couplings between $\Phi$ and $X$ and between $\Phi$ 
and $H$ are allowed.  A simple way of realizing this is to assume 
that $H$ and $X$ are localized in different points in extra 
dimensions, while $\Phi$ have broad wavefunctions overlapping 
with both $H$ and $X$.%
\footnote{Precisely speaking, in extra dimensional theories it is 
 sufficient to assume that the Yukawa couplings are allowed only 
 in places separated from the $X$ field.  The Higgs fields can be 
 delocalized in that case.}

\begin{figure}
\begin{center}
  \center{\includegraphics[width=.49\textwidth]{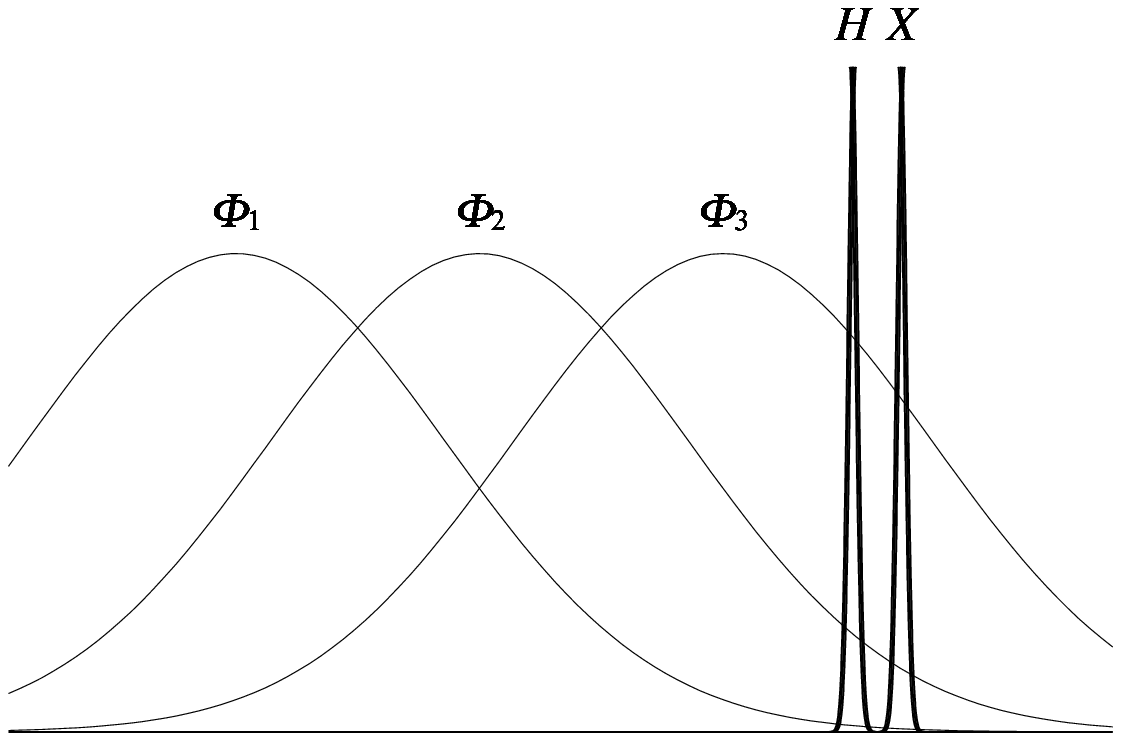}
\hspace{0.1cm}
          \includegraphics[width=.49\textwidth]{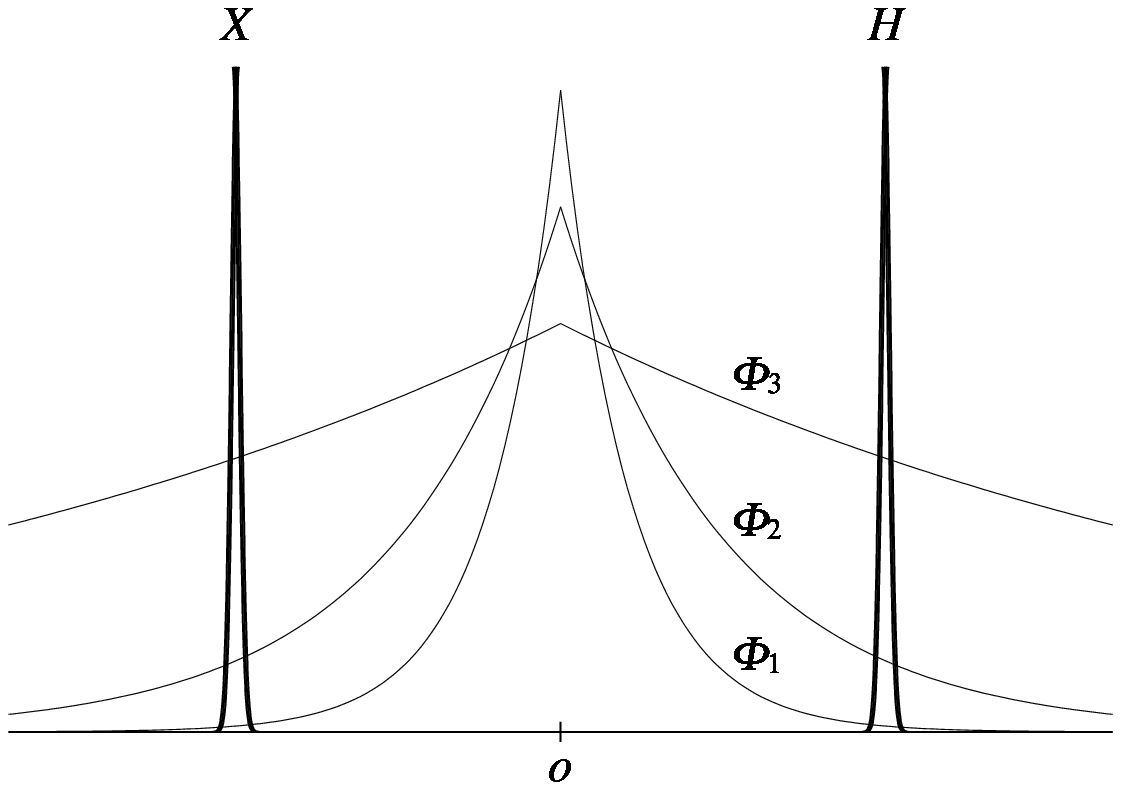}}
\caption{A schematic depiction of possible configurations of the matter, 
 Higgs and supersymmetry breaking fields.  The Higgs and supersymmetry 
 breaking fields are localized to separate but nearby points (left). 
 In the case where all the matter wavefunctions are spherically symmetric 
 and centered around the same point $o$, the Higgs and supersymmetry 
 breaking fields can be localized (approximately) the same distance 
 away from $o$ (right).  The gauge fields are assumed to propagate 
 in the bulk.}
\label{fig:Higgsphobic}
\end{center}
\end{figure}

Let us now focus on the extra dimensional way of suppressing 
${\cal O}_{\zeta_f}$ described above.  In theories with extra dimensions, 
wavefunction overlaps between $\Phi$ and $H$ control the size of the 
4D Yukawa couplings~\cite{Hamidi:1986vh,ArkaniHamed:1999dc,Kaplan:2000av}. 
This motivates a configuration where heavier generation matter fields 
have larger wavefunction overlaps with the Higgs fields.  For example, 
if the widths of all the matter wavefunctions are the same, then 
heavier generation fields have wavefunctions peaked closer to the 
Higgs fields.  On the other hand, if all the matter wavefunctions 
are peaked at the same point (separated from where the Higgs fields 
reside), then heavier generation fields have wider wavefunctions. 
A schematic depiction of these possibilities is shown in 
Fig.~\ref{fig:Higgsphobic}.

The location of $X$ cannot be arbitrary.  If the $X$ field is 
localized to a generic point in the region where matter wavefunctions 
are significant, the generated soft supersymmetry breaking parameters 
have a random structure, e.g. Eq.~(\ref{eq:generic-SUSY}), leading 
to large flavor and $CP$ violation.  One way of avoiding this 
is to localize $X$ far away from wavefunction peaks for all the 
matter fields, in which case flavor violation in the low energy 
supersymmetry breaking masses arises only from loop effects across 
the bulk~\cite{Kaplan:2000av}.  Another way is to localize $X$ at a 
point close to where the Higgs fields are localized~\cite{Nomura:2008pt}. 
In this case the tree-level structure of the operators 
${\cal O}_{\kappa_\Phi,\eta_\Phi}$ is correlated with that 
of the Yukawa couplings ${\cal O}_{\lambda_f}$, since they 
are both controlled by the wavefunction values of the matter 
fields in the region where the $X$ and the Higgs fields reside. 
In the case where all the matter wavefunctions are spherically 
symmetric and peaked at the same point $o$, a similar correlation 
can be obtained by localizing $X$ to a point (approximately) the 
same distance away from $o$ as the Higgs fields.  In all these cases, 
flavor violation also arises from loop effects across the bulk.

Some of the field configurations discussed above are depicted 
schematically in Fig.~\ref{fig:Higgsphobic}.  Note that while 
the figures describe only three matter wavefunctions for illustrative 
purposes, all $Q_i$, $U_i$, $D_i$, $L_i$ and $E_i$ fields can have 
distinct wavefunctions.  The geometry of the extra dimensions can 
also be more general: the number of extra dimensions is arbitrary, 
and the spacetime need not be flat.  While the scale of the extra 
dimensions, i.e. the scale of Kaluza-Klein resonances, is, in principle, 
arbitrary, it is simplest to consider it to be of order the unification 
scale, $1/R \approx M_{\rm unif} \approx 10^{16}~{\rm GeV}$, to preserve 
the success of supersymmetric gauge coupling unification in the most 
straightforward manner.  In the case where $H$ and $X$ are localized 
in the infrared region of warped spacetime, the scale can be lower, 
$10~{\rm TeV} \simlt 1/R \simlt M_{\rm unif}$ (in which case gauge 
coupling unification can occur through modified gauge coupling 
running above $1/R$, as in Ref.~\cite{Gherghetta:2004sq}.)  It is 
also possible to consider the framework in the context of grand 
unification in higher dimensions~\cite{Kawamura:2000ev,Nomura:2006pn}. 
In the theories considered here, a natural scale for flavor physics 
is of order $1/R$, while a natural scale for supersymmetry breaking 
mediation is of order the (local) cutoff scale $M_*$, which we take 
to be somewhat above $1/R$.

One consequence of Higgsphobic supersymmetry breaking is that 
the Higgs sector operators ${\cal O}_{\kappa_H,\eta_H,\mu,b}$ in 
Eqs.~(\ref{eq:Higgs-sector-2}, \ref{eq:Higgs-sector-3}) are forbidden 
in the minimal setup.  There are, however, several ways to generate 
the desired $\mu$ and $b$ parameters, which are discussed in 
Appendix~\ref{app:mu-b_Higgsphobic}.

\subsection{Framework II --- Remote flavor-supersymmetry breaking}
\label{subsec:framework-2}

We now consider the second framework.  An essential ingredient of 
this framework is a ``separation'' between supersymmetry breaking and 
flavor symmetry breaking.  Consider a flavor symmetry $G_{\rm flavor}$ 
that prohibits the Yukawa operators ${\cal O}_{\lambda_f}$ in 
the unbroken limit.  The SSM Yukawa couplings are then generated 
through breaking of $G_{\rm flavor}$, which we assume to be the 
origin of the observed Yukawa structure~\cite{Froggatt:1978nt}. 
An important point is that among the operators relevant to flavor 
violation, ${\cal O}_{Z_\Phi,\kappa_\Phi,\eta_\Phi,\lambda_f,\zeta_f}$ 
in Eqs.~(\ref{eq:matter-sector-1}~--~\ref{eq:Yukawa-sector}), 
${\cal O}_{\zeta_f}$ are the only operators that {\it require} 
both $G_{\rm flavor}$ breaking and supersymmetry breaking (see 
Table~\ref{table:ops}).
\begin{table}
\begin{center}
\begin{tabular}{|c|ccccc|}
\hline
  operators & ${\cal O}_{Z_\Phi}$ & ${\cal O}_{\kappa_\Phi}$ 
    & ${\cal O}_{\eta_\Phi}$ & ${\cal O}_{\lambda_f}$ 
    & ${\cal O}_{\zeta_f}$ \\ \hline
  $G_{\rm flavor}$ breaking & &         &         & $\surd$ & $\surd$ \\
  supersymmetry breaking    & & $\surd$ & $\surd$ &         & $\surd$ \\
\hline
\end{tabular}
\end{center}
\caption{Required symmetry breaking to write down operators in 
 Eqs.~(\ref{eq:matter-sector-1}~--~\ref{eq:Yukawa-sector}).  The 
 operators ${\cal O}_{\lambda_f,\zeta_f}$ require $G_{\rm flavor}$ 
 breaking, while ${\cal O}_{\kappa_\Phi,\eta_\Phi,\zeta_f}$ require 
 supersymmetry breaking.}
\label{table:ops}
\end{table}
Therefore, if we assume that the theory possesses $G_{\rm flavor}$, 
and that $G_{\rm flavor}$ and supersymmetry are broken in different 
sectors of the theory that do not directly communicate with each 
other, a framework which we call {\it remote flavor-supersymmetry 
breaking}, then the operators ${\cal O}_{\zeta_f}$ are absent.

In the present framework, the Yukawa couplings are generated through 
breaking of $G_{\rm flavor}$.  Assuming that the breaking is caused 
by the VEV of a chiral superfield $\phi$, the relevant operators are 
written schematically as
\begin{equation}
  {\cal L} \approx \int\!d^2\theta\, 
    \sum_{i,j} \left( \frac{\phi}{M_*} \right)^{(n_f)_{ij}}
    \Phi_{Li} \Phi_{Rj} H + {\rm h.c.},
\label{eq:Yukawa_remote}
\end{equation}
where $M_*$ is the (effective) cutoff scale and $(n_f)_{ij}$ are integers. 
In general, these operators could generate dangerous scalar trilinear 
interactions through supersymmetry breaking.  Here we assume that 
they do not generate significant scalar trilinear interactions. 
The conditions under which this is indeed the case are discussed in 
Appendix~\ref{app:a_remote} for the general case that $G_{\rm flavor}$ 
is broken by the VEVs of several fields $\phi_m$ ($m = 1,2,\cdots$). 

A symmetry group $G_{\rm flavor}$ needs to be chosen to avoid all 
the low energy flavor and $CP$ violating constraints.  Suppose that 
$G_{\rm flavor}$ were a simple $U(1)$ symmetry, whose breaking controls 
the size of the Yukawa couplings.  In this case the Cabibbo angle, 
$\theta_C$, would be reproduced by the difference of the $U(1)$ charges 
of $Q_1$ and $Q_2$, $q_{Q_1}$ and $q_{Q_2}$, as $\sin\theta_C \approx 
\epsilon^{q_{Q_1}-q_{Q_2}}$, where $\epsilon$ is the dimensionless 
$U(1)$ breaking parameter normalized to have a charge of $-1$, and 
$q_{Q_1} > q_{Q_2}$.  This, however, would lead to too large flavor 
violation in ${\cal O}_{\kappa_Q}$, giving $(\delta^d_{LL})_{12} 
\approx (\delta^d_{LL})_{21} \approx O(\epsilon^{q_{Q_1}-q_{Q_2}}) 
\approx O(\sin\theta_C)$, which needs to be smaller than of order 
$10^{-2}(m_C/600~{\rm GeV})$ to avoid the bound from $\epsilon_K$. 
(Here we have assumed $\eta_A^2 \approx \kappa_\Phi$ and a generic 
Yukawa structure.)  Similar conflicts between the Yukawa structure 
and flavor violating processes also arise in other places.  One 
possibility of avoiding these bounds is to consider more elaborate 
Abelian charge assignments, for example, under $G_{\rm flavor} 
= U(1) \times U(1)$ (see e.g.~\cite{Feng:2007ke}).  Another, perhaps 
simpler, approach is to consider a non-Abelian $G_{\rm flavor}$ 
symmetry under which (at least) the first two generations of quarks 
and leptons having the same standard model gauge quantum numbers 
are in a single $G_{\rm flavor}$ multiplet.  This makes the relevant 
coefficients of ${\cal O}_{\kappa_\Phi}$ proportional to the unit 
matrix, significantly reducing the problem.

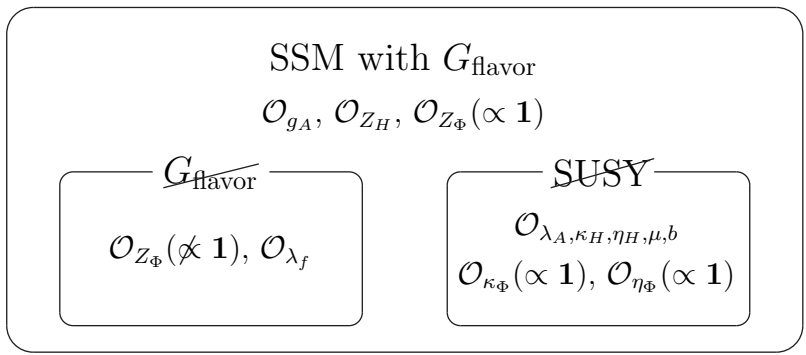
\begin{figure}[t]
\begin{center} 
\begin{picture}(300,140)(0,30)
  \CArc(10,50)(10,180,270)  \CArc(290,50)(10,270,360)
  \CArc(10,160)(10,90,180)  \CArc(290,160)(10,0,90)
  \Line(10,40)(290,40)  \Line(10,170)(290,170)
  \Line(0,50)(0,160)  \Line(300,50)(300,160)
  \Text(150,151)[]{\large SSM with $G_{\rm flavor}$}
  \Text(150,130)[]{${\cal O}_{g_A}$, ${\cal O}_{Z_H}$, 
    ${\cal O}_{Z_\Phi} (\propto {\bf 1})$}
  \CArc(25,55)(5,180,270)  \CArc(129,55)(5,270,360)
  \CArc(25,103)(5,90,180)  \CArc(129,103)(5,0,90)
  \Line(25,50)(129,50)  \Line(25,108)(55,108) \Line(99,108)(129,108)
  \Line(20,55)(20,103)  \Line(134,55)(134,103)
  \Text(77,108)[]{\large $G_{\rm flavor}$} \Line(59,102)(95,111)
  \Text(77,78)[]{${\cal O}_{Z_\Phi} (\npropto {\bf 1})$, 
    ${\cal O}_{\lambda_f}$}
  \CArc(171,55)(5,180,270)  \CArc(275,55)(5,270,360)
  \CArc(171,103)(5,90,180)  \CArc(275,103)(5,0,90)
  \Line(171,50)(275,50)  \Line(171,108)(202,108) \Line(244,108)(275,108)
  \Line(166,55)(166,103)  \Line(280,55)(280,103)
  \Text(225,108)[]{\large SUSY} \Line(205,102)(244,113)
  \Text(223,87)[]{${\cal O}_{\lambda_A, \kappa_H, \eta_H, \mu, b}$}
  \Text(223,69)[]{${\cal O}_{\kappa_\Phi} (\propto {\bf 1})$, 
    ${\cal O}_{\eta_\Phi} (\propto {\bf 1})$}
\end{picture}
\caption{The schematic picture of a remote flavor-supersymmetry breaking 
 theory with $G_{\rm flavor}$ being a sufficiently large subgroup of 
 $SU(3)^5$.  Here, we have depicted only operators relevant for the 
 analysis.}
\label{fig:remote}
\end{center}
\end{figure}

Note that flavor violation in this framework can come mainly from the 
operators ${\cal O}_{Z_\Phi}$ (in the $G_{\rm flavor}$ symmetric field 
basis).  Consider that $G_{\rm flavor}$ is a sufficiently large subgroup 
of $SU(3)^5 = SU(3)_Q \times SU(3)_U \times SU(3)_D \times SU(3)_L 
\times SU(3)_E$ so that all the three generations are treated 
equally under $G_{\rm flavor}$.  In this case the sector breaking 
supersymmetry generates the operators ${\cal O}_{\kappa_\Phi}$ 
and ${\cal O}_{\eta_\Phi}$, but they are completely flavor universal:
\begin{equation}
  (\kappa_\Phi)_{ij} \propto \delta_{ij},
\qquad
  (\eta_\Phi)_{ij} \propto \delta_{ij}.
\label{eq:SU3-1}
\end{equation}
Flavor violation, however, still arises at $M_F$ because the operators 
${\cal O}_{Z_\Phi}$ receive flavor nonuniversal contributions from the 
sector breaking $G_{\rm flavor}$:
\begin{equation}
  (Z_\Phi)_{ij} \npropto \delta_{ij}.
\label{eq:SU3-2}
\end{equation}
This situation is depicted schematically in Fig.~\ref{fig:remote}. 
The scalar squared masses and trilinear interactions in the basis 
where the fields are canonically normalized are, therefore, flavor 
nonuniversal at $M_F$.

A simple way of realizing the present framework is to consider 
higher dimensional theories in which the bulk flavor symmetry 
$G_{\rm flavor}$ and supersymmetry are broken on separate branes. 
Note that $G_{\rm flavor}$ can be broken on multiple branes, which 
could help address the issue of vacuum alignment, depending on 
$G_{\rm flavor}$ and its breaking pattern.  If the relevant extra 
dimension is warped~\cite{Randall:1999ee}, $G_{\rm flavor}$ and 
supersymmetry can be broken at the ultraviolet and infrared branes, 
respectively.  Through the AdS/CFT correspondence, these theories 
have a 4D interpretation that supersymmetry is dynamically broken 
by strong gauge dynamics that have an approximate flavor symmetry. 
Exchanging the locations of supersymmetry and flavor breaking is 
also an interesting possibility, which corresponds to 4D theories 
in which nontrivial flavor structures arise dynamically at low 
energies.

\subsection{Framework III --- Charged supersymmetry breaking}
\label{subsec:framework-3}

The final framework we consider is one in which $X$ carries 
a nontrivial charge of some symmetry, so that the operators 
${\cal O}_{\zeta_f}$ are forbidden.  (We assume that the Yukawa 
couplings, ${\cal O}_{\lambda_f}$, are allowed.)  This symmetry should 
have anomalies with respect to the standard model gauge group so that 
the gaugino mass operators ${\cal O}_{\lambda_A}$ can be written. 
(For an example of this class of models, see~\cite{Nomura:2008pt}.) 
An immediate consequence of this framework, which we call {\it charged 
supersymmetry breaking}, is that the operators ${\cal O}_{\eta_H}$ 
and ${\cal O}_{\eta_\Phi}$ are also forbidden.  This class of 
theories, therefore, has vanishing scalar trilinear interactions 
at the scale $M$.%
\footnote{It is possible that the mechanism generating 
 ${\cal O}_{\lambda_A} \approx \int\!d^2\theta\, (\ln X) 
 {\cal W}^{A\alpha} {\cal W}^A_\alpha + {\rm h.c.}$ also generates 
 other operators, e.g. ${\cal O}_{\eta_H}$ and ${\cal O}_{\eta_\Phi}$ 
 of the form $\int\!d^4\theta\, \ln(X^\dagger X) H^\dagger H$ 
 and $\int\!d^4\theta\, \ln(X^\dagger X) \Phi_i^\dagger \Phi_j 
 + {\rm h.c.}$, of similar size.  In the minimal case such as 
 the one in Ref.~\cite{Nomura:2008pt}, ${\cal O}_{\lambda_A}$ are 
 generated by gauge mediation and the other operators are suppressed 
 (except for ${\cal O}_{\kappa_H}$ and ${\cal O}_{\kappa_\Phi}$ 
 with $(\kappa_\Phi)_{ij} \propto (Z_\Phi)_{ij}$).  The scales of 
 the $M_A$ and the other supersymmetry breaking masses are comparable 
 if $\langle X \rangle \approx g_A^2 M/16\pi^2$.  We here assume this 
 structure: ${\cal O}_{\eta_H,\eta_\Phi}$ are absent, and the $M_A$ 
 are of the same order as the characteristic scale of the other 
 supersymmetry breaking masses.  If the symmetry is nonlinearly 
 realized on $X$ at $M$, the operators ${\cal O}_{\eta_\Phi} \approx 
 \int\!d^4\theta\, (X+X^\dagger) \Phi_i^\dagger \Phi_j + {\rm h.c.}$ 
 are generically expected, leading to a spectrum similar to the 
 case discussed in section~\ref{subsec:framework-1}.}

The operator ${\cal O}_\mu$, which leads to the $\mu$ parameter, may 
or may not be forbidden, depending on the charge assignments of $X$ 
and $H$.  An interesting point is that once we choose the charge 
assignment such that ${\cal O}_\mu$ is allowed, ${\cal O}_b$ is 
always forbidden.  Assuming that the gravitino mass is small, $m_{3/2} 
\ll m_{\rm C,N}$, this implies that $|b| \ll |\mu|^2$, solving the 
supersymmetric $CP$ problem associated with the Higgs sector.  In the 
case where ${\cal O}_\mu$ is not allowed, the $\mu$ and $b$ parameters 
can be generated from ${\cal O}_{\rm SUGRA}$, as long as $m_{3/2}$ 
is of order the weak scale.

The framework of charged supersymmetry breaking is compatible with 
many theories of flavor, including theories with extra dimensions or 
flavor symmetries.  For example, it can be combined with the framework 
described in the previous subsection.  This will prohibit the operators 
${\cal O}_{\eta_H,\eta_\Phi}$, which would otherwise be there. 
Another interesting way of obtaining the Yukawa hierarchy in this 
framework is to generate it at lower energies by some strong gauge 
dynamics~\cite{Nelson:2000sn}.  This generates large anomalous 
dimensions for $\Phi$, and, after canonically normalizing fields, 
the Yukawa and supersymmetry breaking parameters develop a hierarchy. 
Note that unlike Higgsphobic or remote flavor-supersymmetry breaking, 
charged supersymmetry breaking guarantees that the scalar trilinear 
interactions vanish at $M$, which is necessary to prevent reintroducing 
the superpotential flavor problem in models of the type given in 
Ref.~\cite{Nelson:2000sn} (although the problem may be avoided by 
making the gauginos much heavier than the scalars, as discussed in 
(iii) in section~\ref{subsec:general-sol}).  In these theories, the 
scale for supersymmetry breaking mediation can be high to naturally 
preserve gauge coupling unification, e.g. $M \simgt M_{\rm unif}$, 
while the scale of flavor physics, $M_F$, can be much lower as long 
as the mechanism generating the Yukawa structure does not introduce 
large relative running between the standard model gauge couplings. 
In these models, the scale of flavor physics can be as low as 
$10$\,--\,$100~{\rm TeV}$.

\section{Superparticle Spectra and Low Energy Constraints}
\label{sec:spectra}

In the previous section we have presented three classes of theories in 
which ${\cal O}_{\zeta_f}$ are naturally suppressed.  This, however, 
is not sufficient to avoid all the flavor and $CP$ constraints while 
keeping superparticles light.  The constraints from general left-right 
sfermion propagation discussed in section~\ref{subsec:general-LR}, 
as well as those from left-left and right-right sfermion propagation, 
must still be addressed.  The strongest constraints on left-left and 
right-right sfermion propagation arise from $\epsilon_K$ and the 
$\mu \rightarrow e \gamma$ process, giving
\begin{equation}
  \sqrt{|{\rm Im}(\delta^d_{LL/RR})_{12}^2|} 
    \simlt 1 \times 10^{-2} \left( \frac{m_{\rm C}}{600~{\rm GeV}} \right),
\qquad
  \sqrt{|{\rm Im}(\delta^d_{LL})_{12}(\delta^d_{RR})_{12}|} 
    \simlt 2 \times 10^{-4} \left( \frac{m_{\rm C}}{600~{\rm GeV}} \right),
\label{eq:LL-RR_eK}
\end{equation}
\begin{equation}
  |(\delta^e_{LL})_{12}| \simlt 6 \times 10^{-4} \frac{10}{\tan\beta} 
    \left( \frac{m_{\rm N}}{200~{\rm GeV}} \right)^2,
\qquad
  |(\delta^e_{RR})_{12}| \simlt 3 \times 10^{-3} \frac{10}{\tan\beta} 
    \left( \frac{m_{\rm N}}{200~{\rm GeV}} \right)^2.
\label{eq:LL-RR_mu-e-gamma}
\end{equation}
Here, the bounds are obtained conservatively by scanning the 
ratios of the superparticle masses in the same range as that 
leading to Eqs.~(\ref{eq:mu-e-gamma}~--~\ref{eq:Hg-EDM}) (see 
e.g.~\cite{Gabbiani:1996hi}).%
\footnote{Here, we have also scanned the region $0.3\, \alpha_1/\alpha_2 
 \simlt M_1^2/M_2^2 \simlt 3\, \alpha_1/\alpha_2$ and $1 \simlt 
 \mu^2/m_{\rm N}^2 \simlt 16$, and required that $10\%$ of the region 
 evades the experimental constraints.  If we change these conditions, 
 the bounds would change by a factor of a few, but our conclusions 
 would be unaffected.}
The bounds from $\epsilon_K$ are obtained from the conservative 
requirement that the supersymmetric contribution does not exceed the 
experimental value of $|\epsilon_{K,{\rm exp}}| \simeq 2.23 \times 
10^{-3}$~\cite{Amsler:2008zz}.  The $\mu \rightarrow e \gamma$ 
process also leads to bounds on $|(\delta^\nu_{LL})_{ij}|$ which 
provide similar constraints as the bound on $|(\delta^e_{LL})_{12}|$ 
in the theories considered below.

In this section we perform general analyses on flavor and $CP$ constraints 
in the classes of theories discussed in section~\ref{sec:solutions}, 
assuming that the Yukawa structure coefficients ${\cal E}^f_{ij}$ 
factorize.  This is possible because the supersymmetry breaking 
parameters follow a definite pattern in each class of theories, which, 
up to $O(1)$ coefficients, is described by a few free parameters.

\subsection{Factorized flavor structure}
\label{subsec:factorize}

In many of the theories discussed in section~\ref{sec:solutions}, the 
Yukawa structure coefficients ${\cal E}^f_{ij}$ take a factorized form: 
${\cal E}^u_{ij} = \epsilon_{Q_i} \epsilon_{U_j}$, ${\cal E}^d_{ij} 
= \epsilon_{Q_i} \epsilon_{D_j}$ and ${\cal E}^e_{ij} = \epsilon_{L_i} 
\epsilon_{E_j}$, where $\epsilon_{\Phi_1} \leq \epsilon_{\Phi_2} 
\leq \epsilon_{\Phi_3}$ without loss of generality.  The Yukawa 
couplings, Eq.~(\ref{eq:y_f-E}), are then given by
\begin{equation}
  (y_u)_{ij} \approx \tilde{y}\, \epsilon_{Q_i} \epsilon_{U_j},
\qquad
  (y_d)_{ij} \approx \tilde{y}\, \epsilon_{Q_i} \epsilon_{D_j},
\qquad
  (y_e)_{ij} \approx \tilde{y}\, \epsilon_{L_i} \epsilon_{E_j}.
\label{eq:y_f-epsilon}
\end{equation}
In fact, this factorization generically appears in models with 
Abelian flavor symmetries and those with extra dimensions.  Models 
with non-Abelian symmetries may also obey this, for example, if 
the $SU(3)_\Phi$ symmetry is broken by three spurions, $\approx 
(0,0,\epsilon_{\Phi_3})$, $(0,\epsilon_{\Phi_2},\epsilon_{\Phi_2})$, 
and $(\epsilon_{\Phi_1},\epsilon_{\Phi_1},\epsilon_{\Phi_1})$ for each 
$\Phi = Q,U,D,L,E$.  The Yukawa couplings of Eq.~(\ref{eq:y_f-epsilon}) 
lead to the following quark and lepton masses and mixings
\begin{equation}
\begin{array}{lll}
  (m_t,m_c,m_u) & \approx & \tilde{y}\, \langle H_u \rangle\, 
    (\epsilon_{Q_3}\epsilon_{U_3},\, \epsilon_{Q_2}\epsilon_{U_2},\, 
     \epsilon_{Q_1}\epsilon_{U_1}), \\
  (m_b,m_s,m_d) & \approx & \tilde{y}\, \langle H_d \rangle\, 
    (\epsilon_{Q_3}\epsilon_{D_3},\, \epsilon_{Q_2}\epsilon_{D_2},\, 
     \epsilon_{Q_1}\epsilon_{D_1}), \\
  (m_\tau,m_\mu,m_e) & \approx & \tilde{y}\, \langle H_d \rangle\, 
    (\epsilon_{L_3}\epsilon_{E_3},\, \epsilon_{L_2}\epsilon_{E_2},\, 
     \epsilon_{L_1}\epsilon_{E_1}), \\
  (m_{\nu_\tau},m_{\nu_\mu},m_{\nu_e}) & \approx & 
    \frac{\tilde{y}^2 \langle H_u \rangle^2}{M_N}\, 
    (\epsilon_{L_3}^2,\, \epsilon_{L_2}^2,\, \epsilon_{L_1}^2),
\end{array}
\label{eq:q-l-masses}
\end{equation}
and
\begin{equation}
  V_{\rm CKM} \approx
  \left( \begin{array}{ccc}
    1 & \epsilon_{Q_1}/\epsilon_{Q_2} & \epsilon_{Q_1}/\epsilon_{Q_3} \\
    \epsilon_{Q_1}/\epsilon_{Q_2} & 1 & \epsilon_{Q_2}/\epsilon_{Q_3} \\
    \epsilon_{Q_1}/\epsilon_{Q_3} & \epsilon_{Q_2}/\epsilon_{Q_3} & 1
  \end{array} \right),
\qquad
  V_{\rm MNS} \approx
  \left( \begin{array}{ccc}
    1 & \epsilon_{L_1}/\epsilon_{L_2} & \epsilon_{L_1}/\epsilon_{L_3} \\
    \epsilon_{L_1}/\epsilon_{L_2} & 1 & \epsilon_{L_2}/\epsilon_{L_3} \\
    \epsilon_{L_1}/\epsilon_{L_3} & \epsilon_{L_2}/\epsilon_{L_3} & 1
  \end{array} \right),
\label{eq:q-l-mixings}
\end{equation}
where we have included the neutrino masses through the seesaw 
mechanism with the right-handed neutrino Majorana masses 
$W \approx M_N \epsilon_{N_i} \epsilon_{N_j} N_i N_j$, and 
$V_{\rm CKM}$ and $V_{\rm MNS}$ are the quark and lepton mixing 
matrices, respectively.

The values of the $\epsilon$ parameters are constrained by 
the observed quark and lepton masses and mixings through 
Eqs.~(\ref{eq:q-l-masses}, \ref{eq:q-l-mixings}).  They may also 
be constrained by possible grand unification.  In the analysis 
of this section we use the following values for $\epsilon_{\Phi_i}$, 
inferred from the quark and lepton masses and mixing run up to 
the unification scale~\cite{Fusaoka:1998vc}:
\begin{equation}
\begin{array}{lll}
  \epsilon_{Q_1} \approx 0.003\, 
    \tilde{y}^{-\frac{1}{2}} \alpha_q,      \quad & 
  \epsilon_{U_1} \approx 0.001\, 
    \tilde{y}^{-\frac{1}{2}} \alpha_q^{-1}, \quad & 
  \epsilon_{D_1} \approx 0.002\, 
    \tilde{y}^{-\frac{1}{2}} \alpha_q^{-1} \tan\beta,
\\
  \epsilon_{Q_2} \approx 0.03\, 
    \tilde{y}^{-\frac{1}{2}} \alpha_q,      \quad & 
  \epsilon_{U_2} \approx 0.04\, 
    \tilde{y}^{-\frac{1}{2}} \alpha_q^{-1}, \quad & 
  \epsilon_{D_2} \approx 0.004\, 
    \tilde{y}^{-\frac{1}{2}} \alpha_q^{-1} \tan\beta,
\\
  \epsilon_{Q_3} \approx 0.7\, 
    \tilde{y}^{-\frac{1}{2}} \alpha_q,      \quad & 
  \epsilon_{U_3} \approx 0.7\, 
    \tilde{y}^{-\frac{1}{2}} \alpha_q^{-1}, \quad & 
  \epsilon_{D_3} \approx 0.01\, 
    \tilde{y}^{-\frac{1}{2}} \alpha_q^{-1} \tan\beta,
\end{array}
\label{eq:epsilon-QUD}
\end{equation}
\begin{equation}
\begin{array}{ll}
  \epsilon_{L_1} \approx 0.002\, 
    \tilde{y}^{-\frac{1}{2}} \alpha_l\, \tan\beta, \quad & 
  \epsilon_{E_1} \approx 0.001\, 
    \tilde{y}^{-\frac{1}{2}} \alpha_l^{-1},
\\
  \epsilon_{L_2} \approx 0.008\, 
    \tilde{y}^{-\frac{1}{2}} \alpha_l\, \tan\beta, \quad & 
  \epsilon_{E_2} \approx 0.04\, 
    \tilde{y}^{-\frac{1}{2}} \alpha_l^{-1},
\\
  \epsilon_{L_3} \approx 0.01\, 
    \tilde{y}^{-\frac{1}{2}} \alpha_l\, \tan\beta, \quad & 
  \epsilon_{E_3} \approx 0.7\, 
    \tilde{y}^{-\frac{1}{2}} \alpha_l^{-1},
\end{array}
\label{eq:epsilon-LE}
\end{equation}
where $\alpha_{q,l}$ are numbers parameterizing the freedoms unfixed 
by the data.  We have chosen $\alpha_{q,l}$ so that $SU(5)$ grand 
unified relations, $\epsilon_{Q_i} = \epsilon_{U_i} = \epsilon_{E_i}$ 
and $\epsilon_{D_i} = \epsilon_{L_i}$, are almost satisfied with 
$\alpha_{q} = \alpha_{l} = 1$.  Note that the precise numbers in 
Eqs.~(\ref{eq:epsilon-QUD}, \ref{eq:epsilon-LE}) are not very important 
because of unknown $O(1)$ coefficients in the expressions of the quark 
and lepton masses and mixings as well as the bounds from low energy 
flavor and $CP$ violation.  In addition, while we have used the data 
at $M_F \simeq M_{\rm unif}$, using a lower value of $M_F$ would not 
qualitatively change the results, as it would only change the numbers 
in Eqs.~(\ref{eq:epsilon-QUD}, \ref{eq:epsilon-LE}) by additional 
$O(1)$ factors.

\subsection{Higgsphobic supersymmetry breaking}
\label{subsec:spectra-1}

Higgsphobic supersymmetry breaking theories discussed in 
section~\ref{subsec:framework-1} give the following pattern for 
the flavor supersymmetry breaking parameters at $M_F$:
\begin{equation}
  (m_\Phi^2)_{ij} \approx \{ \epsilon_{\Phi_i} \epsilon_{\Phi_j} 
    + (\eta_\Phi^\dagger \eta_\Phi)_{ij} 
    + \Delta^\Phi_{ij} \}\, m_{\rm S}^2,
\label{eq:mij_Higgsphobic}
\end{equation}
\begin{equation}
  (a_f)_{ij} \approx 
    \{ (y_f)_{kj} (\eta_{\Phi_L})_{ki} + (y_f)_{ik} (\eta_{\Phi_R})_{kj}
    \}\, m_{\rm S},
\label{eq:aij_Higgsphobic}
\end{equation}
where we have suppressed flavor universal contributions as well as 
a possible difference between the colored and non-colored superparticle 
mass scales.  Here, $m_{\rm S}$ is the characteristic scale for 
supersymmetry breaking parameters, $(\eta_{\Phi})_{ij} \approx 
\epsilon_{\Phi_i} \epsilon_{\Phi_j}$ are complex $3 \times 3$ 
matrices, and $\Delta^\Phi_{ij}$ parameterize flavor violating effects 
arising from bulk loops in higher dimensional theories discussed in 
section~\ref{subsec:framework-1}.    In flat space models, we typically 
find $\Delta^\Phi_{ii} \simlt g^2/16\pi^2 \approx O(10^{-2})$ because 
they arise from bulk gauge loops, where $g$ represents the standard 
model gauge couplings.  The off-diagonal components are smaller, 
$\Delta^\Phi_{ij} (i \neq j) \ll \Delta^\Phi_{ii}$, since they arise 
through brane-localized terms which are volume suppressed.  On the 
other hand, in warped space models where $H$ and $X$ are in the 
infrared region, we expect $\Delta^\Phi_{ij} \simlt \epsilon_{\Phi_i} 
\epsilon_{\Phi_j}$, since in the dual 4D picture any couplings of 
a matter field to $H$ and $X$, including flavor violating loops 
in higher dimensions, are controlled by the anomalous dimension of 
the corresponding strong dynamics operator, which provides a factor 
$\epsilon_{\Phi_i}$ for each $\Phi_i$.%
\footnote{In these warped space models, the gaugino masses $M_A$ are 
 likely to be somewhat suppressed compared with $m_{\rm S}$, and a 
 flavor universal contribution to the scalar squared masses at $M_F$, 
 $\delta m_\Phi^2|_{\rm univ} \simlt M_A^2$, is also expected.  Using 
 naive dimensional analysis in the dual 4D picture, we find $M_A 
 \approx g_A^2 (N/16\pi^2)^{3/4} m_{\rm S}/\tilde{y}^{1/2}$, where 
 $N$ is the size of the strongly coupled sector yielding $H$ and 
 $X$.  This does not significantly affect the analysis below in some 
 parameter region, although the top squarks may be somewhat heavy 
 in these theories.}

The pattern of Eqs.~(\ref{eq:mij_Higgsphobic}, \ref{eq:aij_Higgsphobic}) 
is essentially the one discussed in Ref.~\cite{Nomura:2007ap} (for 
$\Delta^\Phi_{ij} \simlt \epsilon_{\Phi_i} \epsilon_{\Phi_j}$).  The 
left-left and right-right mass insertion parameters generated by 
Eq.~(\ref{eq:mij_Higgsphobic}) are
\begin{eqnarray}
  && (\delta^u_{LL})_{ij} \approx 
    \left\{ (1 + \epsilon_{Q_3}^2) \epsilon_{Q_i} \epsilon_{Q_j} 
    + (\Delta^Q_i - \Delta^Q_j) \frac{\epsilon_{Q_i}}{\epsilon_{Q_j}} 
    \right\} \frac{m_{\rm S}^2}{m_{\rm C}^2},
\qquad
  (\delta^u_{RR})_{ij} \approx 
    (\delta^u_{LL})_{ij}\Bigr|_{Q \rightarrow U},
\label{eq:u_LL-RR_Higgsphobic}\\
  && (\delta^d_{LL})_{ij} \approx 
    \left\{ (1 + \epsilon_{Q_3}^2) \epsilon_{Q_i} \epsilon_{Q_j} 
    + (\Delta^Q_i - \Delta^Q_j) \frac{\epsilon_{Q_i}}{\epsilon_{Q_j}} 
    \right\} \frac{m_{\rm S}^2}{m_{\rm C}^2},
\qquad
  (\delta^d_{RR})_{ij} \approx 
    (\delta^d_{LL})_{ij}\Bigr|_{Q \rightarrow D},
\label{eq:d_LL-RR_Higgsphobic}\\
  && (\delta^e_{LL})_{ij} \approx 
    \left\{ (1 + \epsilon_{L_3}^2) \epsilon_{L_i} \epsilon_{L_j} 
    + (\Delta^L_i - \Delta^L_j) \frac{\epsilon_{L_i}}{\epsilon_{L_j}} 
    \right\} \frac{m_{\rm S}^2}{m_{\rm N}^2},
\qquad\,\,\,
  (\delta^e_{RR})_{ij} \approx 
    (\delta^e_{LL})_{ij}\Bigr|_{L \rightarrow E},
\label{eq:e_LL-RR_Higgsphobic}
\end{eqnarray}
for $i < j$, and $(\delta^f_{LL})_{ij} \approx (\delta^f_{LL})_{ji}$ 
and $(\delta^f_{RR})_{ij} \approx (\delta^f_{RR})_{ji}$.  Here, we 
have retained only the diagonal components of $\Delta^\Phi_{ij}$, 
$\Delta^\Phi_i \equiv \Delta^\Phi_{ii}$, which is justified in most 
models, as discussed above.  The left-right mass insertion parameters 
given by Eq.~(\ref{eq:aij_Higgsphobic}) are
\begin{eqnarray}
  && (\delta^u_{LR})_{ij} = (\delta^u_{RL})^*_{ji} \approx 
    \epsilon_{Q_i} \epsilon_{U_j} (\epsilon_{Q_j}^2 + \epsilon_{U_i}^2) 
    \frac{v \sin\beta}{m_{\rm C}},
\label{eq:u_LR-RL_Higgsphobic}\\
  && (\delta^d_{LR})_{ij} = (\delta^d_{RL})^*_{ji} \approx 
    \epsilon_{Q_i} \epsilon_{D_j} (\epsilon_{Q_j}^2 + \epsilon_{D_i}^2) 
    \frac{v \cos\beta}{m_{\rm C}},
\label{eq:d_LR-RL_Higgsphobic}\\
  && (\delta^e_{LR})_{ij} = (\delta^e_{RL})^*_{ji} \approx 
    \epsilon_{L_i} \epsilon_{E_j} (\epsilon_{L_j}^2 + \epsilon_{E_i}^2) 
    \frac{v \cos\beta}{m_{\rm N}},
\label{eq:e_LR-RL_Higgsphobic}
\end{eqnarray}
where we have assumed that the renormalization group effect makes 
$m_{\rm S} \rightarrow m_{\rm C}$ and $m_{\rm N}$ in the scalar 
trilinear interactions for colored and non-colored superparticles, 
respectively.  Note that we have suppressed all the subleading 
terms as well as $O(1)$ coefficients in 
Eqs.~(\ref{eq:u_LL-RR_Higgsphobic}~--~\ref{eq:e_LR-RL_Higgsphobic}).

One finds that $(\delta^f_{LR})_{ij}$ in 
Eqs.~(\ref{eq:u_LR-RL_Higgsphobic}~--~\ref{eq:e_LR-RL_Higgsphobic}) are 
typically much smaller than those in Eq.~(\ref{eq:delta-LR-ij}) with 
$a_{\rm C,N} \approx m_{\rm C,N}$ for $i,j=1,2$, so that the left-right 
flavor violation caused by this source is small.  For the multiple 
mass insertion diagrams discussed in section~\ref{subsec:general-LR}, 
we find that the effective mass insertion parameters of 
Eq.~(\ref{eq:delta-LR_eff}) for $i,j = 1,2$ are given in terms of
\begin{eqnarray}
  d^f_{ij} &\approx& {\rm max}\Biggl\{ 
  (\epsilon_{\Phi_{Lj}}^2+\epsilon_{\Phi_{R3}}^2) 
    \hat{\Delta}^{\Phi_L}_{i3} \frac{c_d m_{\rm S}^2}{m_{\rm C,N}^2},
\,\,\,
  (\epsilon_{\Phi_{L3}}^2+\epsilon_{\Phi_{Ri}}^2) 
    \hat{\Delta}^{\Phi_R}_{3j} \frac{c_d m_{\rm S}^2}{m_{\rm C,N}^2},
\nonumber\\
  && \qquad 
  \hat{\Delta}^{\Phi_L}_{i3} \hat{\Delta}^{\Phi_R}_{3j} 
    \frac{c_t \tilde{y}\, m_{\rm C} m_{\rm S}^4\, t}{m_{\rm C,N}^5},
\,\,\,
  (\epsilon_{\Phi_{L3}}^2+\epsilon_{\Phi_{Ri}}^2) 
    (\epsilon_{\Phi_{Lj}}^2+\epsilon_{\Phi_{R3}}^2) 
    \epsilon_{\Phi_{L3}}^2 \epsilon_{\Phi_{R3}}^2
    \frac{c_t \tilde{y}\, v^2 m_{\rm C}}{m_{\rm C,N}^3\, t} 
  \Biggr\},
\label{eq:d-f_ij}
\end{eqnarray}
as
\begin{equation}
  (\delta^f_{LR,{\rm eff}})_{ij} \approx 
    \epsilon_{\Phi_{Li}} \epsilon_{\Phi_{Rj}} \frac{v}{m_{\rm C,N}\, t} 
  {\rm max}\Biggl\{ d^f_{ij},
\,\,\,
  \hat{\Delta}^{\Phi_L}_{ij} 
    \frac{c_d \tilde{y}\, m_{\rm C} m_{\rm S}^2\, t}{m_{\rm C,N}^3},
\,\,\,
  \hat{\Delta}^{\Phi_R}_{ij} 
    \frac{c_d \tilde{y}\, m_{\rm C} m_{\rm S}^2\, t}{m_{\rm C,N}^3} 
  \Biggr\},
\quad (i \neq j),
\label{eq:Delta_eff-ij}
\end{equation}
and
\begin{equation}
  (\delta^f_{LR,{\rm eff}})_{ii} 
    \approx \epsilon_{\Phi_{Li}} \epsilon_{\Phi_{Ri}} 
    \frac{v}{m_{\rm C,N}\, t}\, d^f_{ii}.
\label{eq:Delta_eff-ii}
\end{equation}
Here,
\begin{eqnarray}
  \hat{\Delta}^{\Phi_L}_{ij} &\equiv& 
    (\Delta^{\Phi_L}_i - \Delta^{\Phi_L}_j)\, \theta_L 
    + \epsilon_{\Phi_{Lj}}^2,
\qquad\,
  \theta_L = \left\{ \begin{array}{ll}
    1 & \mbox{for } i<j \\
    \epsilon_{\Phi_{Lj}}^2/\epsilon_{\Phi_{Li}}^2 & \mbox{for } i>j
  \end{array}, \right.
\label{eq:hat-Delta-L}\\
  \hat{\Delta}^{\Phi_R}_{ij} &\equiv& 
    (\Delta^{\Phi_R}_i - \Delta^{\Phi_R}_j)\, \theta_R 
    + \epsilon_{\Phi_{Ri}}^2,
\qquad
  \theta_R = \left\{ \begin{array}{ll}
    \epsilon_{\Phi_{Ri}}^2/\epsilon_{\Phi_{Rj}}^2 & \mbox{for } i<j \\
    1 & \mbox{for } i>j
  \end{array}, \right.
\label{eq:hat-Delta-R}
\end{eqnarray}
$(\Phi_L,\Phi_R,m_{\rm C,N},t) = (Q,U,m_{\rm C},1), 
(Q,D,m_{\rm C},\tan\beta), (L,E,m_{\rm N},\tan\beta)$ for $f = u,d,e$, 
and we have assumed $\epsilon_{\Phi_1} \simlt \epsilon_{\Phi_2} \simlt 
\epsilon_{\Phi_3} \simlt O(1)$, $\tan\beta \simgt O(1)$, and $\mu 
\approx m_{\rm C}$.  The values of $(\delta^f_{LR,{\rm eff}})_{ij}$ 
above should be compared with the ``naive'' $(\delta^f_{LR})_{ij}$ 
in Eq.~(\ref{eq:delta-LR-ij}) with $a_{\rm C,N} \approx m_{\rm C,N}$:
\begin{equation}
  (\delta^f_{LR,{\rm naive}})_{ij} \approx 
    \epsilon_{\Phi_{Li}} \epsilon_{\Phi_{Ri}} \frac{v}{m_{\rm C,N}\, t}.
\label{eq:delta-LR-naive}
\end{equation}
Using Eqs.~(\ref{eq:epsilon-QUD}, \ref{eq:epsilon-LE}) and $\Delta^\Phi_i 
\simlt {\rm max}\{ g^2/16\pi^2, \epsilon_{\Phi_i}^2 \}$, and neglecting 
contributions sufficiently smaller than $(\delta^f_{LR,{\rm naive}})_{ij}$, 
the expressions of Eqs.~(\ref{eq:Delta_eff-ij}, \ref{eq:Delta_eff-ii}) for 
moderate $\tan\beta \approx O(10)$ and $\tilde{y} \approx O(1)$ become
\begin{equation}
  (\delta^f_{LR,{\rm eff}})_{ij} 
  \approx (\delta^f_{LR,{\rm naive}})_{ij}\, 
  {\rm max}\Biggl\{ 
    (y_f)_{33}^2 \frac{c_d m_{\rm S}^2}{\tilde{y}^2 m_{\rm C,N}^2},
\,\,\,
    (y_f)_{33}^2 
    \frac{c_t m_{\rm C} m_{\rm S}^4\, t}{\tilde{y}\, m_{\rm C,N}^5},
\,\,\,
    (y_f)_{33}^4 
    \frac{c_t v^2 m_{\rm C}}{\tilde{y}^3 m_{\rm C,N}^3\, t} 
  \Biggr\}.
\label{eq:delta-LR-ratio}
\end{equation}
(For larger $\tilde{y}$, we have additional potentially relevant 
contributions of order $\Delta^\Phi_{ij} c_d \tilde{y}\, m_{\rm C} 
m_{\rm S}^2 t/m_{\rm C,N}^3$ inside the curly brackets.)   From this, 
we find that $(\delta^f_{LR,{\rm eff}})_{ij}$ can in fact be smaller 
than $(\delta^f_{LR,{\rm naive}})_{ij}$, so that the supersymmetric 
left-right flavor problem can be solved.  Natural values of 
$(\delta^f_{LR,{\rm eff}})_{ij}$ inferred from Eq.~(\ref{eq:delta-LR-ratio}), 
however, are not very much smaller than $(\delta^f_{LR,{\rm naive}})_{ij}$ 
(typically no more than an order of magnitude for $f=e$), so we can 
still expect positive signatures in future search on flavor and $CP$ 
violation, e.g. in EDM experiments, in this class of theories.  For 
larger $\tan\beta$, it becomes increasingly difficult to obtain 
$(\delta^f_{LR,{\rm eff}})_{ij} \ll (\delta^f_{LR,{\rm naive}})_{ij}$, 
so that very large $\tan\beta$, e.g. $\tan\beta \simgt 30$, is disfavored.

We now consider constraints from left-left and right-right sfermion 
propagation.  Using Eqs.~(\ref{eq:epsilon-QUD}, \ref{eq:epsilon-LE}) 
in Eqs.~(\ref{eq:d_LL-RR_Higgsphobic}, \ref{eq:e_LL-RR_Higgsphobic}), 
the bounds of Eqs.~(\ref{eq:LL-RR_eK}, \ref{eq:LL-RR_mu-e-gamma}) give
\begin{equation}
  (\Delta^Q_1 - \Delta^Q_2) 
    + 9 \times 10^{-4}\, \frac{\alpha_q^2}{\tilde{y}} 
\,\,\simlt\,\, 
  0.1 \left( \frac{m_{\rm C}}{600~{\rm GeV}} \right) 
    \frac{m_{\rm C}^2}{m_{\rm S}^2},
\label{eq:LL-LL-d}
\end{equation}
\begin{equation}
  (\Delta^D_1 - \Delta^D_2) 
    + 2 \times 10^{-5}\, \frac{\tan^2\!\beta}{\tilde{y}\, \alpha_q^2} 
\,\,\simlt\,\, 
  2 \times 10^{-2} \left( \frac{m_{\rm C}}{600~{\rm GeV}} \right) 
    \frac{m_{\rm C}^2}{m_{\rm S}^2},
\label{eq:RR-RR-d}
\end{equation}
\begin{equation}
  \left\{ (\Delta^Q_1 - \Delta^Q_2) 
    + 9 \times 10^{-4}\, \frac{\alpha_q^2}{\tilde{y}} \right\}^{1/2} 
  \left\{ (\Delta^D_1 - \Delta^D_2) 
    + 2 \times 10^{-5}\, \frac{\tan^2\!\beta}{\tilde{y}\, \alpha_q^2} 
  \right\}^{1/2} 
\,\simlt\,\, 
  9 \times 10^{-4} \left( \frac{m_{\rm C}}{600~{\rm GeV}} \right) 
    \frac{m_{\rm C}^2}{m_{\rm S}^2},
\label{eq:LL-RR-d}
\end{equation}
\begin{equation}
  (\Delta^L_1 - \Delta^L_2) 
    + 6 \times 10^{-5}\, \frac{\alpha_l^2 \tan^2\!\beta}{\tilde{y}} 
\,\,\simlt\,\, 
  2 \times 10^{-3} \frac{10}{\tan\beta} 
    \left( \frac{m_{\rm N}}{200~{\rm GeV}} \right)^2 
    \frac{m_{\rm N}^2}{m_{\rm S}^2},
\label{eq:LL-e}
\end{equation}
\begin{equation}
  (\Delta^E_1 - \Delta^E_2) 
    + 2 \times 10^{-3}\, \frac{1}{\tilde{y}\, \alpha_l^2} 
\,\,\simlt\,\, 
  0.1\, \frac{10}{\tan\beta} 
    \left( \frac{m_{\rm N}}{200~{\rm GeV}} \right)^2 
    \frac{m_{\rm N}^2}{m_{\rm S}^2},
\label{eq:RR-e}
\end{equation}
where we have assumed $\epsilon_{\Phi_3} \simlt O(1)$.  While there 
is an $O(1)$ coefficient omitted in front of each term, we still find 
some tension in Eqs.~(\ref{eq:RR-RR-d}, \ref{eq:LL-RR-d}, \ref{eq:LL-e}). 
For $m_{\rm S} \approx m_{\rm N} \approx m_{\rm C}/(2$\,--\,$4)$, for 
example, these bounds require
\begin{equation}
  \frac{1}{\tilde{y}\, \alpha_q^2} 
\,\,\simlt\,\, 
  10^2 \left( \frac{10}{\tan\beta} \right)^2
    \left( \frac{m_{\rm C}}{600~{\rm GeV}} \right),
\label{eq:RR-RR-d-2}
\end{equation}
\begin{equation}
  (\Delta^Q_1 - \Delta^Q_2)^{1/2} 
  \left\{ (\Delta^D_1 - \Delta^D_2) 
    + 2 \times 10^{-5}\, \frac{\tan^2\!\beta}{\tilde{y}\, \alpha_q^2} 
  \right\}^{1/2} 
\,\simlt\,\, 
  10^{-2} \left( \frac{m_{\rm C}}{600~{\rm GeV}} \right),
\label{eq:LL-RR-d-2}
\end{equation}
\begin{equation}
  (\Delta^L_1 - \Delta^L_2) 
    + 6 \times 10^{-5}\, \frac{\alpha_l^2 \tan^2\!\beta}{\tilde{y}} 
\,\,\simlt\,\, 
  10^{-3} \frac{10}{\tan\beta} 
    \left( \frac{m_{\rm N}}{200~{\rm GeV}} \right)^2,
\label{eq:LL-e-2}
\end{equation}
at the order of magnitude level.  The conditions of 
Eqs.~(\ref{eq:RR-RR-d-2}, \ref{eq:LL-RR-d-2}) are satisfied 
in a wide parameter region, while the condition of 
Eq.~(\ref{eq:LL-e-2}) requires
\begin{equation}
  (\Delta^L_1 - \Delta^L_2) 
  \,\,\simlt\,\, 10^{-3} \frac{10}{\tan\beta} 
    \left( \frac{m_{\rm N}}{200~{\rm GeV}} \right)^2,
\qquad
  \frac{\alpha_l^2}{\tilde{y}} 
  \,\,\simlt\,\, 0.1 \left( \frac{10}{\tan\beta} \right)^3 
    \left( \frac{m_{\rm N}}{200~{\rm GeV}} \right)^2,
\label{eq:LL-e-3}
\end{equation}
unless there is a strong cancellation.  The first inequality leads to 
a tension in theories with $\Delta^\Phi_i \approx g^2/16\pi^2 \approx 
O(10^{-2})$, although it can be ameliorated by taking somewhat large 
$m_{\rm N}$, e.g. $m_{\rm N} \simgt 600~{\rm GeV} (\tan\beta/10)^{1/2}$ 
or smaller $\tan\beta$.  On the other hand, theories with $\Delta^\Phi_i 
\approx \epsilon_{\Phi_i}^2$ have little tension, and taking somewhat 
small $\alpha_l$ is enough to avoid the bounds.  Note that very large 
$\tan\beta$ is, again, disfavored.

Finally, we discuss implications on the superparticle spectrum. 
The intergenerational mass splittings between the sfermions are 
controlled by Eq.~(\ref{eq:mij_Higgsphobic}), leading to
\begin{equation}
  |m_{\Phi_i}^2 - m_{\Phi_j}^2| \approx m_{\rm S}^2\,\, 
    {\rm max}\{ \Delta^\Phi_{i,j}, \epsilon_{\Phi_{i,j}}^2 \}.
\label{eq:Higgsphobic_split-1}
\end{equation}
In particular, this gives
\begin{eqnarray}
  |m_{\tilde{\tau}_R}^2 - m_{\tilde{\mu}_R}^2| 
  &\approx& m_{\rm S}^2\,\, {\rm max}\{ \Delta^E_2, \Delta^E_3, 
    \epsilon_{E_3}^2 \},
\label{eq:Higgsphobic_tau-mu}\\
  |m_{\tilde{\tau}_R}^2 - m_{\tilde{e}_R}^2| 
  &\approx& m_{\rm S}^2\,\, {\rm max}\{ \Delta^E_1, \Delta^E_3, 
    \epsilon_{E_3}^2 \},
\label{eq:Higgsphobic_tau-e}\\
  |m_{\tilde{\mu}_R}^2 - m_{\tilde{e}_R}^2| 
  &\approx& m_{\rm S}^2\,\, {\rm max}\{ \Delta^E_1, \Delta^E_2, 
    \epsilon_{E_2}^2 \},
\label{eq:Higgsphobic_mu-e}
\end{eqnarray}
which can lead to $O(1)$ fractional mass splittings between $\tilde{\tau}_R$ 
and $\tilde{e}_R,\tilde{\mu}_R$, and $O(10^{-3}$--$10^{-2})$ fractional 
mass splitting between $\tilde{\mu}_R$ and $\tilde{e}_R$.  The signs 
of the splittings are arbitrary, so that, for example, $\tilde{\tau}_R$ 
can be heavier than $\tilde{e}_R,\tilde{\mu}_R$.  Similar levels of 
intergenerational mass splittings are also possible for other sfermions, 
although for squarks, splittings will be somewhat diluted by large 
flavor universal renormalization effects by a factor of order 
$m_{\rm S}^2/m_{\rm C}^2$.

\subsection{Remote flavor-supersymmetry breaking}
\label{subsec:spectra-2}

The remote flavor-supersymmetry breaking scenario discussed in 
section~\ref{subsec:framework-2} can give a variety of patterns for 
the supersymmetry breaking parameters, depending on $G_{\rm flavor}$ 
and its breaking.  In general, $G_{\rm flavor}$ can be a product of 
a ``3 generation,'' ``2 generation,'' or ``single generation'' symmetry 
acting on each $\Phi = Q,U,D,L,E$.  The first class corresponds to 
(a sufficiently large subgroup of) $SU(3)$ acting on three generations 
$(\Phi_1,\Phi_2,\Phi_3)$, the second to (a sufficiently large subgroup 
of) $SU(2)$ acting on the first two generations $(\Phi_1,\Phi_2)$, and 
the third to products of $U(1)$ symmetries. 

In the limit of unbroken $G_{\rm flavor}$, the coefficients of the 
matter supersymmetry breaking operators are given by
\begin{equation}
  (\kappa_\Phi)_{ij} \approx \delta_{ij} \kappa^\Phi_i,
\qquad
  (\eta_\Phi)_{ij} \approx \delta_{ij} \eta^\Phi_i,
\label{eq:coeff-remote-gen}
\end{equation}
in the field basis where $(Z_\Phi)_{ij} = \delta_{ij}$.  Here, depending 
on the component of $G_{\rm flavor}$ acting on $\Phi_i$, the parameters 
$\kappa^\Phi_i$ and $\eta^\Phi_i$ exhibit the following pattern:
\begin{eqnarray}
  && \kappa^\Phi_1 = \kappa^\Phi_2 = \kappa^\Phi_3,
\qquad
  \eta^\Phi_1 = \eta^\Phi_2 = \eta^\Phi_3,
\qquad
  \mbox{for ``3 generation'' symmetry},
\label{eq:remote_3-gen}\\
  && \kappa^\Phi_1 = \kappa^\Phi_2 \neq \kappa^\Phi_3,
\qquad
  \eta^\Phi_1 = \eta^\Phi_2 \neq \eta^\Phi_3,
\qquad
  \mbox{for ``2 generation'' symmetry},
\label{eq:remote_2-gen}\\
  && \kappa^\Phi_1 \neq \kappa^\Phi_2 \neq \kappa^\Phi_3,
\qquad
  \eta^\Phi_1 \neq \eta^\Phi_2 \neq \eta^\Phi_3,
\qquad
  \mbox{for ``single generation'' symmetry}.
\label{eq:remote_1-gen}
\end{eqnarray}
In any of these theories, our framework provides a solution to the 
superpotential problem.  To see if all the constraints are avoided, 
we also need to study effects coming from $G_{\rm flavor}$ violation.

We now focus on the case where $G_{\rm flavor}$ contains a ``3 generation'' 
symmetry for each $\Phi = Q,U,D,L,E$, e.g. $G_{\rm flavor} = SU(3)^5$. 
We also assume that $G_{\rm flavor}$ is broken by three spurions, $\approx 
(0,0,\epsilon_{\Phi_3})$, $(0,\epsilon_{\Phi_2},\epsilon_{\Phi_2})$, 
and $(\epsilon_{\Phi_1},\epsilon_{\Phi_1},\epsilon_{\Phi_1})$ 
for each $\Phi = Q,U,D,L,E$, which guarantees that ${\cal E}^f_{ij}$ 
take a factorized form.  The operator coefficients at $M_F$ in 
the field basis that naturally realizes $G_{\rm flavor}$ are 
then given by
\begin{equation}
  (Z_\Phi)_{ij} \approx \delta_{ij} 
    + \gamma\, \epsilon_{\Phi_i} \epsilon_{\Phi_j},
\qquad
  (\kappa_\Phi)_{ij} \approx \delta_{ij} + (D\mbox{-term}),
\qquad
  (\eta_\Phi)_{ij} \approx \delta_{ij},
\label{eq:coeff-remote}
\end{equation}
where we have omitted $O(1)$ coefficients that appear in each 
$(i,j)$-element of the second term of $(Z_\Phi)_{ij}$, in front of 
the first term of $(\kappa_\Phi)_{ij}$, and in front of the expression 
for $(\eta_\Phi)_{ij}$.  Here, $\gamma$ parameterizes the strength of 
the $G_{\rm flavor}$ breaking effect, which is suppressed by the volume 
of the extra dimensions in a higher dimensional realization of the 
scenario.  (If the matter fields have nontrivial wavefunctions and/or 
$G_{\rm flavor}$ is broken on several different branes, $\gamma$ 
can depend on $\Phi,i,j$.)  The second term of $(\kappa_\Phi)_{ij}$, 
denoted as $D$-term, arises if $G_{\rm flavor}$ contains a continuous 
gauge symmetry component (for $M_F \simlt M$), but it is absent if 
$G_{\rm flavor}$ is a global or discrete symmetry.

After canonically normalizing fields, $(Z_\Phi)_{ij} = \delta_{ij}$, 
Eq.~(\ref{eq:coeff-remote}) leads to the following operator coefficients:
\begin{equation}
  (\kappa_\Phi)_{ij} \approx \delta_{ij} 
    (1 + \gamma \epsilon_{\Phi_i}^2 + \Delta^\Phi_i),
\qquad
  (\eta_\Phi)_{ij} \approx 
    \delta_{ij} (1 + \gamma \epsilon_{\Phi_i}^2),
\label{eq:coeff-remote-2}
\end{equation}
where we have diagonalized $(Z_\Phi)_{ij}$ of Eq.~(\ref{eq:coeff-remote}) 
and then rescaled fields so that the $(Z_\Phi)_{ij}$ become $\delta_{ij}$. 
Here, the third term of $(\kappa_\Phi)_{ij}$ parameterizes the possible 
$D$-term contribution (which is always flavor diagonal in this basis). 
This contribution arises when a continuous gauged $G_{\rm flavor}$ 
symmetry is broken by the VEVs of fields $\phi_m$ ($m = 1,2,\cdots$) 
which have supersymmetry breaking masses~\cite{Drees:1986vd}.  For 
example, if $G_{\rm flavor}$ is broken by $\phi_1$ and $\phi_2$ whose 
transformation properties under $G_{\rm flavor}$ is opposite, a $D$-term 
contribution of order $m_{\phi_1}^2 - m_{\phi_2}^2$ generically arises, 
where $m_{\phi_m}^2$ represents the supersymmetry breaking mass squared 
of $\phi_m$.  While the $D$-term contribution is generically dangerous 
in theories with a gauged flavor symmetry, in the present framework 
the supersymmetry breaking masses of $\phi_m$ are suppressed because 
of the separation between $G_{\rm flavor}$ and supersymmetry breaking, 
so that the resulting $D$-term contribution $\Delta^\Phi_i$ is also 
suppressed.  (This was observed in Ref.~\cite{Kobayashi:2002mx} in 
theories with $G_{\rm flavor} = U(1)$.)  The contribution is typically 
suppressed by a loop factor, as well as by powers of the ratio of the 
compactification scale, $M_c$, to the (effective) cutoff scale, $M_*$, 
in theories with extra dimensions.  We therefore expect $\Delta^\Phi_i 
\approx (M_c/M_*)^n/16\pi^2 \simlt O(10^{-2})$, where $n$ is a 
model-dependent integer which can in general depend on $\Phi$ and $i$.%
\footnote{If different $\phi_m$ have different renormalizable interactions 
 of order $\lambda$, $\Delta^\Phi_i \approx (\lambda^2/16\pi^2)^2 
 (m_{3/2}/m_{\rm S})^2$ can be generated through anomaly mediation. 
 This contribution is typically of order $10^{-4}$ or smaller.}

Note that in the field basis leading to Eq.~(\ref{eq:coeff-remote-2}) 
the Yukawa couplings are still given by Eq.~(\ref{eq:y_f-epsilon}). 
The left-left and right-right mass insertion parameters are then 
given by
\begin{eqnarray}
  && (\delta^u_{LL})_{ij} \approx 
    \left\{ \gamma (1 + \gamma \epsilon_{Q_3}^2) \epsilon_{Q_i} \epsilon_{Q_j} 
    + (\Delta^Q_i - \Delta^Q_j) \frac{\epsilon_{Q_i}}{\epsilon_{Q_j}} 
    \right\} \frac{m_{\rm S}^2}{m_{\rm C}^2},
\qquad
  (\delta^u_{RR})_{ij} \approx 
    (\delta^u_{LL})_{ij}\Bigr|_{Q \rightarrow U},
\label{eq:u_LL-RR_remote}\\
  && (\delta^d_{LL})_{ij} \approx 
    \left\{ \gamma (1 + \gamma \epsilon_{Q_3}^2) \epsilon_{Q_i} \epsilon_{Q_j} 
    + (\Delta^Q_i - \Delta^Q_j) \frac{\epsilon_{Q_i}}{\epsilon_{Q_j}} 
    \right\} \frac{m_{\rm S}^2}{m_{\rm C}^2},
\qquad
  (\delta^d_{RR})_{ij} \approx 
    (\delta^d_{LL})_{ij}\Bigr|_{Q \rightarrow D},
\label{eq:d_LL-RR_remote}\\
  && (\delta^e_{LL})_{ij} \approx 
    \left\{ \gamma (1 + \gamma \epsilon_{L_3}^2) \epsilon_{L_i} \epsilon_{L_j} 
    + (\Delta^L_i - \Delta^L_j) \frac{\epsilon_{L_i}}{\epsilon_{L_j}} 
    \right\} \frac{m_{\rm S}^2}{m_{\rm N}^2},
\qquad\,\,\,
  (\delta^e_{RR})_{ij} \approx 
    (\delta^e_{LL})_{ij}\Bigr|_{L \rightarrow E},
\label{eq:e_LL-RR_remote}
\end{eqnarray}
for $i < j$, and $(\delta^f_{LL})_{ij} \approx (\delta^f_{LL})_{ji}$ 
and $(\delta^f_{RR})_{ij} \approx (\delta^f_{RR})_{ji}$.  The flavor 
and $CP$ violating left-right mass insertion parameters are given by
\begin{eqnarray}
  && (\delta^u_{LR})_{ij} = (\delta^u_{RL})^*_{ji} \approx \gamma 
    \epsilon_{Q_i} \epsilon_{U_j} (\epsilon_{Q_j}^2 + \epsilon_{U_i}^2) 
    \frac{v \sin\beta}{m_{\rm C}},
\label{eq:u_LR-RL_remote}\\
  && (\delta^d_{LR})_{ij} = (\delta^d_{RL})^*_{ji} \approx \gamma 
    \epsilon_{Q_i} \epsilon_{D_j} (\epsilon_{Q_j}^2 + \epsilon_{D_i}^2) 
    \frac{v \cos\beta}{m_{\rm C}},
\label{eq:d_LR-RL_remote}\\
  && (\delta^e_{LR})_{ij} = (\delta^e_{RL})^*_{ji} \approx \gamma 
    \epsilon_{L_i} \epsilon_{E_j} (\epsilon_{L_j}^2 + \epsilon_{E_i}^2) 
    \frac{v \cos\beta}{m_{\rm N}}.
\label{eq:e_LR-RL_remote}
\end{eqnarray}
We find that the mass insertion parameters of 
Eqs.~(\ref{eq:u_LL-RR_remote}~--~\ref{eq:e_LR-RL_remote}) 
take the same form as those of 
Eqs.~(\ref{eq:u_LL-RR_Higgsphobic}~--~\ref{eq:e_LR-RL_Higgsphobic}) 
for $\gamma = 1$, although the origins of $\Delta^\Phi_i$ are different. 
Since $\gamma$ is expected to be smaller than $1$, the present class of 
theories is at least as safe as Higgsphobic theories with $\Delta^\Phi_i 
\approx O(10^{-2})$ from the flavor and $CP$ violation point of 
view. Moreover, since we expect $\Delta^\Phi_i < O(10^{-2})$ in most 
cases due to power suppression of $(M_c/M_*)^n$, or even absent if 
$G_{\rm flavor}$ is a global or discrete symmetry, the low energy 
constraints from flavor and $CP$ violation generically give little 
tension with LHC observability in the present class of theories.

The intergenerational mass splittings between sfermions have a similar 
formula to the Higgsphobic case:
\begin{equation}
  |m_{\Phi_i}^2 - m_{\Phi_j}^2| \approx m_{\rm S}^2\,\, 
    {\rm max}\{ \Delta^\Phi_{i,j}, \gamma \epsilon_{\Phi_{i,j}}^2 \}.
\label{eq:remote_split-1}
\end{equation}
In particular, the right-handed sleptons have the splittings
\begin{eqnarray}
  |m_{\tilde{\tau}_R}^2 - m_{\tilde{\mu}_R}^2| 
  &\approx& m_{\rm S}^2\,\, {\rm max}\{ \Delta^E_2, \Delta^E_3, 
    \gamma \epsilon_{E_3}^2 \},
\label{eq:remote_tau-mu}\\
  |m_{\tilde{\tau}_R}^2 - m_{\tilde{e}_R}^2| 
  &\approx& m_{\rm S}^2\,\, {\rm max}\{ \Delta^E_1, \Delta^E_3, 
    \gamma \epsilon_{E_3}^2 \},
\label{eq:remote_tau-e}\\
  |m_{\tilde{\mu}_R}^2 - m_{\tilde{e}_R}^2| 
  &\approx& m_{\rm S}^2\,\, {\rm max}\{ \Delta^E_1, \Delta^E_2, 
    \gamma \epsilon_{E_2}^2 \}.
\label{eq:remote_mu-e}
\end{eqnarray}
The size of the splittings depends on the details of the theory, 
specifically the size of $\Delta^E_i$ and $\gamma$.  We expect 
that the fractional mass splittings between $\tilde{\tau}_R$ 
and $\tilde{e}_R,\tilde{\mu}_R$ and between $\tilde{\mu}_R$ and 
$\tilde{e}_R$ are at most of $O(1)$ and $O(10^{-2})$, respectively, 
and typically smaller by at least a factor of a few because of 
the suppression by $\gamma$ and $(M_c/M_*)^n$.

In the above analysis, we have focused on the case that $G_{\rm flavor}$ 
is the product of five ``3 generation'' symmetries acting on $Q,U,D,L,E$. 
Similar analyses, however, can also be performed in the case where 
(some of) $Q,U,D,L,E$ have only ``2 generation'' or ``single generation'' 
symmetries.  In particular, in the case of a ``2 generation'' symmetry, 
we expect that the conclusion on flavor and $CP$ violation does not 
change because the constraints from the processes involving the third 
generation particles are weak.  The fractional mass splittings between 
the third and first two generation sfermions in this case are expected 
to be of $O(1)$, without a suppression from $\gamma$ or $(M_c/M_*)^n$. 
In the case of a ``single generation'' symmetry, model by model 
analyses are needed.  We expect, however, that the model can avoid 
the constraints if it involves ``single generation'' symmetries only 
for some $\Phi$.  For example, we can consider only $E$ has a 
``single generation'' symmetry while the rest have ``3 generation'' 
symmetries, e.g. $G_{\rm flavor} = SU(3)_Q \times SU(3)_U \times 
SU(3)_D \times SU(3)_L \times U(1)_E$.  In this case the fractional 
mass splittings between $\tilde{\tau}_R$ and $\tilde{\mu}_R$ and 
between $\tilde{\mu}_R$ and $\tilde{e}_R$ are of the same order, 
and presumably of $O(1)$.

\subsection{Charged supersymmetry breaking}
\label{subsec:spectra-3}

If the supersymmetry breaking field is charged under some symmetry, the 
operators ${\cal O}_{\eta_\Phi}$ are forbidden.  In fact, this charged 
supersymmetry breaking framework can be combined with many flavor 
theories.  To parameterize these wide classes of theories, we somewhat 
arbitrarily take
\begin{equation}
  (\kappa_\Phi)_{ij} \approx \delta_{ij} 
    + \gamma \epsilon_{\Phi_i} \epsilon_{\Phi_j} 
    + \delta_{ij} \Delta^\Phi_i,
\qquad
  (\eta_\Phi)_{ij} = 0,
\label{eq:coeff-charged}
\end{equation}
at the scale $M_F$.  This parameterization accommodates many theories 
of flavor, including ones based on extra dimensions, strong dynamics, 
and ``3 generation'' flavor symmetries.  (In some of these theories, 
the flavor universal part, i.e. the first term, of $(\kappa_\Phi)_{ij}$
is absent, but this does not affect the analysis of flavor and $CP$ 
violation.)  This parameterization needs to be modified appropriately 
for other types of theories, for example those based on ``2 generation'' 
or ``single generation'' flavor symmetries.

The left-left and right-right mass insertion parameters generated by 
Eq.~(\ref{eq:coeff-charged}) are
\begin{eqnarray}
  && (\delta^u_{LL})_{ij} \approx 
    \left\{ \gamma \epsilon_{Q_i} \epsilon_{Q_j} 
    + (\Delta^Q_i - \Delta^Q_j) \frac{\epsilon_{Q_i}}{\epsilon_{Q_j}} 
    \right\} \frac{m_{\rm S}^2}{m_{\rm C}^2},
\qquad
  (\delta^u_{RR})_{ij} \approx 
    (\delta^u_{LL})_{ij}\Bigr|_{Q \rightarrow U},
\label{eq:u_LL-RR_charged}\\
  && (\delta^d_{LL})_{ij} \approx 
    \left\{ \gamma \epsilon_{Q_i} \epsilon_{Q_j} 
    + (\Delta^Q_i - \Delta^Q_j) \frac{\epsilon_{Q_i}}{\epsilon_{Q_j}} 
    \right\} \frac{m_{\rm S}^2}{m_{\rm C}^2},
\qquad
  (\delta^d_{RR})_{ij} \approx 
    (\delta^d_{LL})_{ij}\Bigr|_{Q \rightarrow D},
\label{eq:d_LL-RR_charged}\\
  && (\delta^e_{LL})_{ij} \approx 
    \left\{ \gamma \epsilon_{L_i} \epsilon_{L_j} 
    + (\Delta^L_i - \Delta^L_j) \frac{\epsilon_{L_i}}{\epsilon_{L_j}} 
    \right\} \frac{m_{\rm S}^2}{m_{\rm N}^2},
\qquad\,\,\,
  (\delta^e_{RR})_{ij} \approx 
    (\delta^e_{LL})_{ij}\Bigr|_{L \rightarrow E},
\label{eq:e_LL-RR_charged}
\end{eqnarray}
for $i < j$, and $(\delta^f_{LL})_{ij} \approx (\delta^f_{LL})_{ji}$ 
and $(\delta^f_{RR})_{ij} \approx (\delta^f_{RR})_{ji}$, while the 
flavor and $CP$ violating left-right mass insertion parameters are
\begin{eqnarray}
  && (\delta^f_{LR})_{ij} = (\delta^f_{RL})^*_{ji} \approx 0.
\label{eq:f_LR-RL_charged}
\end{eqnarray}
The constraints from low energy flavor and $CP$ violation are obviously 
not stronger than in the classes of theories discussed in the previous 
two subsections for the same values of $\gamma$ and $\Delta^\Phi_i$. 
Note that while the flavor and $CP$ violating left-right mass insertion 
parameters are vanishing (up to the higher order effects from the Yukawa 
couplings), we still have flavor and $CP$ violating effects generated 
by multiple mass insertion diagrams through the flavor universal 
part of $(\delta^f_{LR})_{ij}$ and through flavor and $CP$ violating 
$(\delta^f_{LL,RR})_{ij}$.  Nontrivial flavor and $CP$ violation, 
therefore, can still be discovered in future experiments such as 
ones in Refs.~[\ref{MEG:X}~--~\ref{Semertzidis:2003iq:X}].

The intergenerational mass splittings between sfermions are given 
by the formula in Eq.~(\ref{eq:remote_split-1}).  The size of the 
splittings is controlled by the parameters $\gamma$ and $\Delta^\Phi_i$, 
which are model dependent.  In many models, $\gamma \simlt O(1)$ and 
$\Delta^\Phi_i \simlt O(10^{-2})$.  We can, however, still expect 
that a variety of patterns for the intergenerational mass splittings 
can be obtained in this framework.

\section{Probing the Origin of Flavor at the LHC}
\label{sec:LHC}

We have seen that supersymmetry with a flavorful spectrum is consistent 
with bounds from low energy flavor and $CP$ violating processes in 
a wide variety of models where the superpotential flavor problem is 
solved.  Furthermore, the spectrum in these models can easily be light 
enough that it will have a substantial production cross section at 
the LHC.  While it appears that the deviation from flavor universality 
is small, especially in the first two generations, the splittings are 
large enough that they can be significant at the LHC.  Consider the 
right-handed sleptons, which from the structure of the SSM renormalization 
group equations are expected to be the lightest sfermions.  If flavor 
nonuniversality comes only from renormalization group evolution, then 
the fractional mass splitting between $\tilde{e}_R$ and $\tilde{\mu}_R$ 
is controlled by the muon Yukawa coupling, and is expected to be of 
$O(10^{-4})$ for $\tan\beta \approx O(10)$.  On the other hand, the 
analysis in section~\ref{sec:spectra} shows that the fractional mass 
splitting from the contribution at $M_F$ can be easily of $O(10^{-2})$. 
The splittings between $\tilde{\tau}_R$ and $\tilde{e}_R,\tilde{\mu}_R$ 
will also be larger than the renormalization group prediction in the 
moderate $\tan\beta$ regime.  Furthermore, the contribution from 
renormalization group flow has a definite sign, with the mass ordering 
of the superparticles being anticorrelated with that of the standard 
model particles, i.e. $m_{\tilde{\tau}_R} < m_{\tilde{\mu}_R} < 
m_{\tilde{e}_R}$.  On the other hand, the contributions considered 
in section~\ref{sec:spectra} could be positive or negative, and could 
thus produce a spectrum which is unambiguously different from flavor 
universality.

An interesting possibility considered in 
Refs.~\cite{Nomura:2007ap,Feng:2007ke} is the case where the lightest 
supersymmetric particle (LSP) is the gravitino, and the next-to-lightest 
supersymmetric particle (NLSP) is one of the right-handed sleptons. 
In the case where the supersymmetry breaking scale is large, i.e. $M$ 
is large, the lifetime of the NLSP is long enough that it escapes 
the detector.  A slepton NLSP will then appear as a heavy charged stable 
particle, something quite spectacular at the LHC.  (Here we consider 
the case where $R$ parity is conserved so that decay of the NLSP into 
only standard model particles is forbidden.)  Furthermore, the LHC 
will usually produce squarks or gluinos, so NLSPs will generally 
be produced after a chain of decays.  This cascade decay must produce 
a slepton in conjunction with a lepton, and because the NLSP is 
right-handed, the coupling to neutrinos is strongly suppressed. 
Therefore, NLSPs will be produced mostly with charged leptons that 
can be used to measure the flavor content of the NLSP.

If we could observe the decay of the NLSP, we could measure the flavor 
content of the NLSP more easily.  In particular, we could precisely 
determine, by observation of flavor violating decays, if the flavor 
eigenbasis differs from the mass eigenbasis for the sleptons.  Flavorful 
models including those discussed in section~\ref{sec:solutions} 
generically have this property, so this is an interesting test of 
intrinsic flavor nonuniversality.  One way to measure decays of 
long-lived NLSP sleptons is to build a stopper detector outside one 
of the main LHC detectors, which would stop some of the NLSPs and 
measure their decays~\cite{Hamaguchi:2004df}.  If the NLSP has 
a sufficiently long lifetime and a large number of decays are observed 
in a stopper detector, then the flavor composition of the NLSP can 
be measured very accurately.

While the scenario with a weak scale gravitino LSP has an NLSP with 
lifetime much longer than the flight time in the detector, the scenario 
can be extended to a much lighter gravitino LSP or another extremely 
weakly interacting particle, such as the axino.  In fact, if the NLSP has 
a lifetime of order $c\tau \approx O(100~{\rm \mu m}$\,--\,$10~{\rm m})$, 
then there is a clean signal of a non-minimum ionizing track with 
a kink that turns into a minimum ionizing track~\cite{Dimopoulos:1996vz}. 
Furthermore, an ATLAS study with the NLSP decaying to photons showed 
that a substantial number of decays can be measured if $c\tau \simlt 
100~{\rm km}$~\cite{ATLAS-TDR}, so it is conceivable that the decay 
of an NLSP can be observed in the detector for a large range of NLSP 
lifetimes.  This kind of measurement would allow us to study the decays 
of the NLSP and gain knowledge on its flavor content.

We now focus on the case where the lightest neutralino is lighter than 
all of the right-handed sleptons, although the analysis also applies 
if all the superparticles promptly decay to a different particle which 
escapes the detector.  In this case, the events are similar to well 
studied supersymmetric missing energy events, but many interesting studies 
of flavor violation can still be done.  For example, if $m_{\tilde{\tau}_R} 
> m_{\chi^0_2} > m_{\tilde{e}_R,\tilde{\mu}_R}$, a natural spectrum in 
the models of section~\ref{sec:solutions} given the $O(1)$ mass splitting 
of the $\tilde{\tau}_R$ from the other sleptons, then the detectors 
can measure a fractional mass splitting between $\tilde{e}_R$ and 
$\tilde{\mu}_R$ as small as $O(10^{-4})$~\cite{Allanach:2008ib}. 
This may help discriminate different classes of theories discussed in 
section~\ref{sec:solutions}.  These and other studies could be performed 
not only in the right-handed slepton sector, but also with squarks 
and left-handed sleptons, which may determine whether the different 
SSM fields transform under different classes of flavor symmetries, 
as discussed in section~\ref{subsec:spectra-2}.  With a little luck, 
the LHC could discover a smoking gun for intrinsic flavor nonuniversal 
effects in the supersymmetry breaking sector, which may hint at 
the high energy theory that gives rise to the standard model flavor 
structure.

\section{Discussion and Conclusions}
\label{sec:concl}

The problem of excessive flavor and $CP$ violation arising from 
generic weak scale supersymmetry breaking parameters has been a guiding 
principle in searching for viable supersymmetric theories.  In particular, 
this has been a strong motivation behind the search for flavor universal 
mediation mechanisms of supersymmetry breaking.  On the other hand, 
the puzzle of flavor already exists in the standard model, and it is 
possible that the mechanism producing the observed Yukawa structure is 
also responsible for the suppression of possibly large flavor and $CP$ 
violation in supersymmetric theories.  How natural is this possibility? 
Is there any generic tension between the constraints from low energy 
flavor and $CP$ violation and the observability of superparticles at 
the LHC, even if we take into account the possibility of a correlation 
between the structures of the Yukawa couplings and supersymmetry 
breaking parameters?

In this paper we have studied, in a model independent way, the question 
of compatibility between the low energy flavor and $CP$ constraints 
and the observability of a nontrivial flavor structure in superparticle 
spectra at the LHC.  We have seen that there is a model independent 
tension arising from the superpotential operators ${\cal O}_{\zeta_f}$ 
leading to scalar trilinear interactions.  In particular, the constraint 
from the $\mu \rightarrow e\gamma$ process pushes the mass scale for 
non-colored superparticles $m_{\rm N}$ relatively high.  Under the 
assumption of a factorized flavor structure, $m_{\rm N}$ should be 
of order a TeV or larger for a natural size of ${\cal O}_{\zeta_f}$. 
Assuming the usual hierarchy between colored and non-colored superparticle 
masses, this pushes up the masses of colored superparticles beyond 
several TeV, making supersymmetry unobservable at the LHC.  Similar, 
though somewhat weaker constraints also arise from the bounds on 
the electron, neutron and mercury EDMs.  In fact, these observables 
also constrain flavor violation arising from other operators 
through multiple mass insertion diagrams.

We have discussed several ways in which these stringent constraints 
are naturally avoided.  They include relaxing the mass hierarchy between 
colored and non-colored superparticles, making the fundamental strength 
of the Yukawa couplings strong, and making the gaugino masses larger 
than the scalar masses.  We have also presented simple frameworks 
in which the dangerous operators ${\cal O}_{\zeta_f}$ are naturally 
absent.  Since these operators are special, they can be absent in the 
low energy effective theory.  We have considered separating the Higgs 
and supersymmetry breaking fields (Higgsphobic), separating supersymmetry 
and flavor symmetry breaking (remote flavor-supersymmetry breaking), 
and assigning a nontrivial charge to the supersymmetry breaking 
field (charged supersymmetry breaking).  These frameworks can be 
combined with a variety of flavor theories, including ones with 
(flat or warped) extra dimensions, strong dynamics, or flavor symmetries. 
In fact, we can consider many variations of flavor models using 
the basic setups discussed in this paper.  The mediation scale of 
supersymmetry breaking, $M$, and the scale for flavor physics, $M_F$, 
can vary by many orders of magnitude in these theories.

We have performed detailed analyses on the constraints from low energy 
flavor and $CP$ violation in the frameworks described above.  We have 
shown that the constraints, including ones arising from multiple mass 
insertion diagrams and left-left and right-right sfermion propagation, 
can be avoided in natural parameter regions while keeping the 
superparticles light.  Expected sizes of flavor and $CP$ violation, 
however, are not too much smaller than the current bounds, so signatures 
in future search on flavor and $CP$ violation are not eliminated. 
The intergenerational mass splittings among sfermions in these theories 
can show a variety of patterns depending on the underlying mechanism 
responsible for the structure of the Yukawa couplings.  For example, 
if SSM multiplets belonging to different representations of the standard 
model gauge group have different flavor symmetry structures, then it 
will show up in the spectrum of superparticles.  In general, it is 
significant that the spectrum of superparticles contains information 
on left-handed and right-handed fields separately, while the Yukawa 
couplings contain only ``products'' of them.

While the size of the fractional mass splittings directly generated 
by the physics of flavor at $M_F$ is not necessarily very large, e.g. 
of $O(10^{-2})$ or smaller for the first two generations, they are 
large enough to significantly affect the phenomenology at the LHC. 
This is because the intergenerational mass splittings generated by 
the standard SSM renormalization group evolution are typically very 
small, so that additional mass splittings can give large effects on 
the structure of the intergenerational superparticle spectrum.  Moreover, 
because the signs of the intergenerational mass splittings caused 
by effects at $M_F$ are arbitrary (at least from the low energy 
effective field theory point of view), these additional splittings 
can change the mass ordering among different generation sfermions. 
In particular, this can make a third generation sfermion heavier than 
the corresponding first two generation sfermions, which can drastically 
affect the signatures at the LHC.  We find it very possible that the 
intergenerational mass splittings of the size implied by the classes 
of theories discussed in this paper will be measured at the LHC.

The LHC will start running this year, and it is expected to give 
us meaningful data on TeV scale physics as early as next year.  If 
supersymmetry is found, it will not only provide an explanation for 
the stability of the gauge hierarchy and a potential dark matter 
candidate, but it will also allow for a substantial number of new 
flavor measurements.  While there are many viable supersymmetric 
models which are flavor universal, we have shown that there are 
also many nonuniversal models which avoid the stringent low energy 
constraints.  If supersymmetry is in fact flavorful, then its 
discovery at the LHC could shed new light on the longstanding 
mystery of the flavor pattern in the standard model.

\section*{Acknowledgments}

This work was supported in part by the U.S. DOE under Contract 
DE-AC02-05CH11231, and in part by the NSF under grant PHY-04-57315. 
The work of Y.N. was also supported by the NSF under grant PHY-0555661, 
by a DOE OJI, and by the Alfred P. Sloan Research Foundation.  The 
work of D.S. was supported by the Alcatel-Lucent Foundation.

\appendix

\section{{\boldmath $\mu$} and {\boldmath $b$} in Higgsphobic 
 Supersymmetry Breaking}
\label{app:mu-b_Higgsphobic}

In minimal Higgsphobic supersymmetry breaking, 
the operators ${\cal O}_{\kappa_H,\eta_H,\mu,b}$ in 
Eqs.~(\ref{eq:Higgs-sector-2}, \ref{eq:Higgs-sector-3}) 
are forbidden.  The $\mu$ and $b$ parameters are then generated 
only by the operator ${\cal O}_{\rm SUGRA}$ through supergravity 
effects~\cite{Giudice:1988yz}, giving
\begin{equation}
  \mu = \frac{\lambda_H m_{3/2}^*}{(Z_{H_u} Z_{H_d})^{1/2}},
\qquad
  b = \frac{\lambda_H |m_{3/2}|^2}{(Z_{H_u} Z_{H_d})^{1/2}},
\label{eq:mu-b}
\end{equation}
where $m_{3/2}$ is the gravitino mass.  The non-holomorphic 
supersymmetry breaking masses are vanishing, $m_{H_u}^2 = m_{H_d}^2 = 0$, 
at the scale $M$.  The expressions of Eq.~(\ref{eq:mu-b}) imply that 
the gravitino mass should be of order the weak scale to use this 
contribution.  The $\mu$ and $b$ parameters of Eq.~(\ref{eq:mu-b}) 
also lead to dangerous $CP$ violation at low energies unless 
${\rm arg}(m_{3/2}) \simeq {\rm arg}(M_A)$, providing an additional 
constraint on the setup.  In the context of the higher dimensional 
models of section~\ref{subsec:framework-1}, a weak scale gravitino 
mass is obtained if $M_* \approx M_{\rm Pl}$ or if there is an 
additional supersymmetry breaking field $X'$ that does not couple 
to the SSM field and has $F_{X'} \approx F_X (M_{\rm Pl}/M_*)$.

It is possible to extend the minimal Higgsphobic setup by introducing 
fields $B$ which directly couple with both $H$ and $X$.  In higher 
dimensional models, these $B$ fields propagate in the bulk, and integrating 
them out can generate the operators ${\cal O}_{\kappa_H,\eta_H,\mu,b}$ 
in the low energy effective field theory below $1/R$.  The operators 
${\cal O}_{\zeta_f}$ can still be absent by arranging the interactions 
of $B$ appropriately, for example by suppressing the couplings of $B$ 
with matter fields.  With these extensions, the generated Higgs sector 
parameters need not take the form in the minimal setup.  In particular, 
the gravitino mass need not be of order the weak scale, and its phase 
need not be aligned with that of $M_A$.  The requirement from suppressing 
$CP$ violation, instead, constrains the representations and interactions 
of the $B$ fields.  For example, if the exchange of $B$ generates $\mu$ 
but not $b$, and the contribution from Eq.~(\ref{eq:mu-b}) is negligible, 
then the problem of $CP$ violation disappears.

\section{{\boldmath $(a_f)_{ij}$} in Remote Flavor-Supersymmetry Breaking}
\label{app:a_remote}

In remote flavor-supersymmetry breaking, the Yukawa couplings are 
generated through breaking of $G_{\rm flavor}$.  Suppose that the 
breaking is caused by the VEVs of several chiral superfields $\phi_m$ 
($m = 1,2,\cdots$).  The Yukawa couplings are then generated from 
operators of the form
\begin{equation}
  {\cal L} = \int\!d^2\theta\, 
    \sum_{i,j} \sum_{\{ (n_f)^m_{ij} \}}\!\! c_{\{ (n_f)^m_{ij} \}} 
    \frac{\prod_m \phi_m^{(n_f)^m_{ij}}}{(M_* C)^{(n_f)_{ij}}}\, 
    \Phi_{Li} \Phi_{Rj} H + {\rm h.c.},
\label{eq:Yukawa-a-gen}
\end{equation}
as
\begin{equation}
  (y_f)_{ij} = \sum_{\{ (n_f)^m_{ij} \}}\!\! c_{\{ (n_f)^m_{ij} \}} 
    \frac{\prod_m \phi_{m,0}^{(n_f)^m_{ij}}}{M_*^{(n_f)_{ij}}},
\label{eq:Yukawa-gen}
\end{equation}
where $\Phi_{Li}$, $\Phi_{Rj}$, $H$ and $\phi_m$ are canonically 
normalized, $M_*$ is the (effective) cutoff scale, $(n_f)^m_{ij}$ 
are integers with $(n_f)_{ij} \equiv \sum_m (n_f)^m_{ij}$, 
and $\phi_{m,0}$ is the lowest component VEV of $\phi_m$. 
The sum $\sum_{\{ (n_f)^m_{ij} \}}$ runs over all possible 
choices of integers $(n_f)^m_{ij}$ consistent with $G_{\rm flavor}$ 
invariance, and $c_{\{ (n_f)^m_{ij} \}}$ are $O(1)$ coefficients 
in front of each term.  Here, we have included the chiral 
compensator field $C = 1 + \theta^2 m_{3/2}$ which encodes 
supergravity effects~\cite{Cremmer:1978hn}.

The operators of Eq.~(\ref{eq:Yukawa-a-gen}) may generate dangerous 
scalar trilinear interactions.  These are given by
\begin{equation}
  (a_f)_{ij} = \sum_{\{ (n_f)^m_{ij} \}}\!\! c_{\{ (n_f)^m_{ij} \}} 
    \frac{\prod_m \phi_{m,0}^{(n_f)^m_{ij}}}{M_*^{(n_f)_{ij}}} 
    \left\{ (n_f)_{ij} m_{3/2} - \sum_{m'} (n_f)^{m'}_{ij} 
    \frac{F_{\phi_{m'}}}{\phi_{m',0}} \right\},
\label{eq:a-gen}
\end{equation}
where $F_{\phi_m}$ is the $F$-term VEV of $\phi_m$.  This shows 
that even with $F_{\phi_m} = 0$, the scalar trilinear interactions 
are generated, which for $m_{3/2} \approx O(m_{\rm C},m_{\rm N})$ 
lead to $(a_f)_{ij} \approx (y_f)_{ij} m_{\rm C,N}$, and are thus 
dangerous~\cite{Ross:2002mr}.%
\footnote{Our language here is different from that used in 
 Ref.~\cite{Ross:2002mr}, in which the $F$-term VEV of a field 
 is defined including a supergravity contribution so that the 
 effect described here is viewed as arising from the $F$-term 
 VEVs of the fields $\phi_m$.}
More generally, if some of the $\phi_m$ are stabilized using 
supersymmetry breaking effects (e.g. if these fields are flat directions 
lifted by higher dimension operators), we obtain $F_{\phi_m}/\phi_{m,0} 
\approx O({\rm max}\{ m_{3/2}, m_\phi \})$ with $m_\phi$ being 
the scale for the supersymmetry breaking masses of $\phi_m$, and 
we obtain a contribution of order $(a_f)_{ij} \approx (y_f)_{ij} 
{\rm max}\{ m_{3/2}, m_\phi \}$.

The contribution to $(a_f)_{ij}$ described above, however, is suppressed 
if one of the following conditions is satisfied:%
\footnote{While completing this paper, Ref.~\cite{Antusch:2008jf} appeared, 
 which discusses the issue considered in this Appendix.  Their main 
 solution corresponds to our (c) below.  They also discuss the case (b).}
\begin{enumerate}
\item[(a)]
The gravitino mass and the $F$-term VEVs for $\phi_m$ are all small, 
$m_a \equiv {\rm max}\{ m_{3/2}, F_{\phi_m}/\phi_{m,0} \} \ll m_{\rm C,N}$. 
In this case, the effect of Eq.~(\ref{eq:a-gen}) is suppressed by a 
factor of $m_a/m_{\rm C,N}$, giving $(\delta^{u,d}_{LR})_{ij} \approx 
(M_{u,d})_{ij} m_a/m_{\rm C}^2$ and $(\delta^e_{LR})_{ij} \approx 
(M_e)_{ij} m_a/m_{\rm N}^2$.
\item[(b)]
The dimensions of the operators in Eq.~(\ref{eq:Yukawa-a-gen}) are 
the same for all $i,j = 1,2,3$, i.e. $(n_f)_{ij} = n_f$, for $f = u,d,e$, 
and $F_{\phi_m}/\phi_{m,0} \ll m_{{\rm C,N}}$ with ${\rm arg}(m_{3/2}) 
= {\rm arg}(M_A)$ (or $(n_f)_{ij} = n_f$ and $F_{\phi_m}/\phi_{m,0}$ 
are nearly equal with their phases aligned with those of $m_{3/2}, M_A$). 
In this case, $(a_f)_{ij}$ is almost proportional to $(y_f)_{ij}$ as 
a matrix, giving a negligible contribution to $(\delta^f_{LR})_{ij}$.
\item[(c)]
The VEVs of $\phi_m$ are stabilized in the supersymmetric limit. 
In this case, we obtain $\langle \phi_m \rangle = \phi_{m,0} C$, 
since any supersymmetric mass must be accompanied with $C$, leading 
to $F_{\phi_m}/\phi_{m,0} = m_{3/2}$.  Equation~(\ref{eq:a-gen}) 
then gives $(a_f)_{ij} = 0$, and the effect disappears.%
\footnote{There is still an effect from the chiral compensator field 
 at a loop level (anomaly mediation)~\cite{Randall:1998uk}.  This effect, 
 however, does not lead to flavor or $CP$ violation at a dangerous 
 level.  There could also be higher order corrections suppressed by 
 powers of $m_\phi/\phi_{m,0}$, where $m_\phi$ represents generic 
 supersymmetry breaking masses in the $G_{\rm flavor}$ breaking 
 sector.  These corrections are also negligible unless the scale 
 of flavor physics is close to the TeV scale.}
\end{enumerate}
We find it simplest to stabilize $\phi_m$ supersymmetrically, (c), 
although we also leave the possibility open to (a) or (b).  (In fact, 
the experimental bounds may be avoided with one of the above conditions 
satisfied only for $i,j = 1,2$.)  Note that the consideration here 
applies to any field that appears in front of $\Phi_{Li} \Phi_{Rj} H$ 
in the superpotential, and whose lowest component VEV gives 
a significant contribution to the Yukawa couplings.

The issue of scalar trilinear interactions generated by the $F$-term 
VEVs of $C$ and the fields appearing in front of $\Phi_{Li} \Phi_{Rj} H$, 
in fact, exists in wider classes of theories.  For example, in theories 
where the hierarchical Yukawa couplings are generated by wavefunction 
profiles of the matter and Higgs fields in extra dimensions, including 
ones discussed in section~\ref{subsec:framework-1}, there are generally 
moduli fields appearing in front of $\Phi_{Li} \Phi_{Rj} H$.  These 
moduli fields must satisfy condition (a) or (c) (option (b) is typically 
not available).  We assume that one of these conditions is satisfied 
when discussing the classes of theories in section~\ref{sec:solutions}.

\end{document}